# Liquid Epitaxial Growth of Two-dimensional Non-layer structured hybrid Perovskite

**Authors:** Yu Bai[1,3†], Haixin Zhang[1†], Mingjing Zhang[1†], Di Wang[1, 5*], Hui Zeng[2*], Hao Xue[1], Guozheng Wu[1], Ying Xie[4], Yuxia Zhang[4], Hao Jing[5], Jing Su[3], Haohai Yu[4], Zhanggui Hu[1], Ruwen Peng[5], Mu Wang[5], Yicheng Wu[1]

**Affiliations:**

[1] Institute of Functional Crystals, Tianjing University of Technology, Tianjing 300384, China.

[2] School of Electronic and Optical Engineering, Nanjing University of Science & Technology, Nanjing 210094, China.

[3] School of Physics & Optoelectronic Engineering, Nanjing University of Information Science & Technology, Nanjing 210044, China.

[4] State Key Laboratory of Crystal materials and Institute of Crystal Materials, Shandong University, Jinan 250100, China.

[5] National Laboratory of Solid State Microstructure, and School of Physics, and Collaborative Innovation Center of Advanced Microstructures. Nanjing University, Nanjing 210093, China.

* Corresponding author. Email: diwang@tjut.edu.cn (D. W.); zenghui@njust.edu.cn (H. Z.).

† These authors contributed equally to this work.

**Abstract:** The hybrid perovskite $CH_3NH_3PbX_3$ (X= Cl, Br, I) is a promising material for developing novel optoelectronic devices. Owing to the intrinsic non-layer structure, it remains challenging to synthesize molecularly thin $CH_3NH_3PbX_3$ with large size. Here, we report a low-cost and highly efficient fabrication route to obtain large-scale single-crystalline 2D $CH_3NH_3PbX_3$ perovskites on a mica substrate via liquid epitaxy. The 2D perovskite is characterized as 8 nm in thickness and hundreds of micrometers in lateral size. First-principles calculations suggest the strong potassium-halogen interactions at the perovskite/mica interface lower the interface energy of perovskites, driving their fast in-plane growth. Spectroscopic investigations reveal 2D $CH_3NH_3PbBr_3$ possess small exciton binding energy of 30 meV, allowing a superior visible-light photodetector with a photoresponsivity of 126 A/W and a bandwidth exceeded 80 kHz. These features demonstrate that liquid epitaxy is a bottom-up approach to fabricate the non-layer structured 2D perovskites, which offer a new material platform for the device applications and fundamental investigations.

**One Sentence Summary:** A low-cost and high-efficiency route to fabricate molecularly thin perovskite film has been discovered, which makes superior new optoelectronic device very promising.

## INTRODUCTION

Organic-inorganic hybrid perovskites (OIHPs) have emerged and attracted tremendous attentions in the opto-electronic applications recently. The advance of low-cost solution fabrication combined with excellent physics properties make OIHPs a promising material for cheap and high performance opto-electronic devices.(*1,2*) The certified power conversion efficiency (PCE) value of OIHPs has exceeded 23.2% at present,(*3*) approaching the commercial silicon solar cells. The materials are composed of several earth-abundant elements and can be fabricated via various low-cost methods.(*4*) Therefore, it is considered OIHPs to be a low-cost alternative to silicon solar cells, but also to play a key role in next-generation batteries, laser, high-energy radiation detector and much more devices.(*5-7*) Apart from OIHPs, two-dimensional (2D) materials are another rapid expanding field in materials science and condensed state physics. The atomic-thickness crystals, such as graphene, transition-metal dichalcogenides (TMDs), phosphorene and hexagonal boron nitride (*h*-BN), exhibit exotic properties different from their bulk counterparts.(*8-11*) However, until now there is no member in current 2D materials has excellent opt-electronic properties similar with the OIHPs. Although monolayer $MoS_2$ is a direct bandgap semiconductor,(*12*) its performance for optoelectronic application is actually limited by weak light absorption and huge exciton binding energy ($E_B$), as well as fast carrier recombination.

Recent advances in perovskites have suggested that the two-dimensional OIHP is a suitable candidate for filling this gap of 2D-materials family in optoelectronics.(*13*) According to the crystalline structures, 2D OIHPs can be classified into two types. One kind is the non-layer structured 2D OIHPs, derived from the bulk counterparts with the 3D framework of $[PbX_6]^{4-}$ octahedra by reducing one dimension down; the other is the layer structured OIHPs (adopting the Ruddlesden-Popper type)(*14, 15*), in which the $[PbX_6]^{4-}$ framework is split into the layer structure by the organic long-chained cations. The inorganic $[PbX_6]^{4-}$ layers are sandwiched by organic long-chained cation layers and interacted with the adjacent layer via the van der Waals force, moreover, the $[PbX_6]^{4-}$ layer thickness could be controlled via the organic chain length. However, the organic chains in layer structured OIHPs also induce some unfavorable characteristics: (i) the dielectric confinement effect arose from the low dielectric constant of organic chains enhances the exciton $E_B$, which is the disadvantage for the photovoltage electronics;(*16*) (ii) the organic spacer decreases the carrier mobility, cause charge accumulation and radiative recombination losses;(*17*) and (iii) the long chain molecules confine the crystalline size, thus it is hard to synthesize 2D OIHPs films with large size.

The non-layer structured OIHPs, possessing small $E_B$ and long carrier diffuse length, are considered to be the most promising material for photovoltage applications,(*18*) while the intrinsic non-layer structure poses a challenge for 2D growth. Here, we report a low-cost wet-chemical strategy to achieve liquid epitaxial growth of 2D non-layer structured OIHPs on a mica substrate. Oleic acid (OA) is involved as the inhibitor to confine the crystal growth at the molecular level; on the other hand, the strong ions interactions between the potassium (K) and the halogens lower the mica/perovskite interface energy and promote 2D epitaxial growth of perovskites on mica. Experimental characterizations have demonstrated that the grown 2D non-layered $MAPbX_3$ ($MA=CH_3NH_3^+$, X= $Cl^-$, $Br^-$, $I^-$) have thicknesses of 8 nm or below, and their lateral sizes exceed hundred micrometers. Furthermore, the density function theory (DFT) calculations reveal that the K/Br interactions induce trap states at the mica/$MAPbBr_3$ interface, which could capture the photocarriers and prolong the carrier lifetime. The photodetectors based on the 2D $MAPbBr_3$ exhibit extraordinary performances.

## RESULTS AND DISCUSSION

### Crystal growth of 2D OIHPs on mica

In order to grow molecular-thickness perovskites, the critical factor is to achieve the anisotropy crystal growth, i.e. promoting the in-plane growth and suppressing the out-of-plane one. In this paper, we take 2D $MAPbBr_3$ as an example to illustrate the liquid epitaxy growth technology and properties of 2D non-layered perovskites, and the corresponding results of $MAPbCl_3$, $MAPbI_3$ and $CsPbBr_3$ are also available in supplementary materials (SM). The 2D $MAPbX_3$ growth procedure is designed and illustrated in Figure 1A. The precursor solution was dipped into a slit channel, which was composed using two stacking fresh cleaved mica. The growth system was then fixed on the spinner base and rotated at 4000 rpm for 30 s to remove the excess solution. After that, the system was placed onto a hot plate and heated at the crystalline temperatures for several hours. Since the vertical growth of perovskites was restricted by the mica pieces, with the solvent evaporation, 2D perovskite sheets grown slowly on mica surfaces. Owing to the incompressibility of liquid makes, it was hard to reduce the solution spacer. In experiments, the perovskite thicknesses grown under above physical-confinement were limited to be several ten nanometers (fig. S1). In order to further reduce the thickness, the perovskite growth was controlled at the molecular level by surface modification, which was usually applied in the synthesis of 2D Nano-crystals (NCs). (*19*) Here, moderate OA was added into the precursor solution as the surface modifier, which would attach on the perovskite surface and inhibit the thickness increase via a chemical-confinement (Fig. 1B).

Combining with the physical- and chemical-confinement technology, the single-crystal 2D $MAPbBr_3$ perovskites, whose thickness thinner than 10 nm and the lateral size about several hundred micrometers, were successfully grown on mica substrate (Fig. 1C-E, and fig. S2). Other 2D non-layered perovskites were also obtained, including $MAPbCl_3$ and $MAPbI_3$, as well as the inorganic perovskite $CsPbBr_3$ (fig. S4-S6). Although OA played an important role in suppressing the thickness expanding of 2D OIHPs, the excess OA would residue and erode the perovskite surface. To obtain smooth surface, the synthesized samples were washed carefully by using cyclohexane. The detailed growth procedure can be found in the supplementary text. (*20*)

Selective-area electron diffraction (SAED) and grazing incidence X-ray diffraction (GIXRD) had demonstrated the phase purity and well-define crystalline nature of the grown 2D $MAPbBr_3$ perovskite (Fig. 1 F and G). The diffraction peaks (Fig. 1G) could be indexed as (0,0,1), (0,0,2), (0,0,3) and (0,0,4) of the cubic structured crystal,(*21*) which suggested the grown 2D $MAPbBr_3$ perovskites were highly *c*-axis-oriented. The rocking curve of the (002) peak (fig. S7) had the full width at half maximum (FWHM) of $0.154^{\circ}$. This value was larger than that of bulk $MAPbBr_3$ crystal,(*21*) which suggested the high surface-to-volume ratio and structure relaxation of 2D perovskites may lower the crystalline quality. The chemical composition of the 2D $MAPbBr_3$ perovskites was analyzed by X-ray photoelectron spectroscopy (XPS, fig. S8). Excess carbon ion signal indicated the residual OA molecules on the perovskite or substrate surface. The ratio of Pb and Br ions was about 1:3.2 which consisted to the theoretical value of $MAPbBr_3$.

### Growth mechanism of 2D OIHPs on mica

To understand the growth mechanisms of 2D $MAPbBr_3$ perovskites on the mica substrate, we designed two experiments to investigate the factors affecting the liquid epitaxy growth. The first experiment was the epitaxy growth of all inorganic $CsPbBr_3$ perovskites on mica. This was aimed to determine whether the organic molecules, i.e. the methyl ammonium (MA) ions, were essential for the epitaxy growth. Because mica was a typical van der Waals crystal as an ideal substrate for 2D material growth,(*22*) whether the driving forces of the $MAPbX_3$ epitaxy growth was the van der Waals interactions between the MA molecules and the mica? We experimentally confirmed that all inorganic 2D $CsPbBr_3$ perovskite without the MA molecules also could be epitaxially grown on mica. The grown $CsPbBr_3$ perovskite had the thickness of 14 nm and size of tens micrometers (fig. S6), indicating the failure of the hypothesis of Van der Waals force dependency. Meanwhile, the experimental observation has verified that the liquid epitaxy growth works as well for all inorganic perovskite.

The second experiment was the epitaxy growth of 2D $MAPbBr_3$ on various substrates, such as silicon, quartz glass, ITO glass and so on, which was devoted to the understanding of the role of substrate to the epitaxy growth. It was found that $MAPbBr_3$ perovskites preferred to crystallize in solution on the silicon, quartz glass and ITO glass surfaces, rather than epitaxially grown (video S1). This observation was in sharp contrast to the epitaxy growth on mica substrate. We attempted to compress the solution spacer, via improving the solution wetting of the substrates using the surface modifier, to confine the crystal floating and growing up. The thicknesses of the grown $MAPbBr_3$ perovskites on the silicon, quartz glass and ITO glass substrate could be reduced to several hundred micrometers (fig. S9). However, this value was still much larger than the 2D perovskites on mica.

Based on two experiments shown above, we proposed that the liquid epitaxy growth of 2D $MAPbBr_3$ perovskites was determined by the interface interactions between the inorganic components of the perovskite and the mica. Density functional theory (DFT) calculations were carried out to further examine the interface binding. Figure 2A illustrates the interface structures before and after the lattice relaxation. The DFT results clearly shown the facts that: (1) The adsorption of K ion was exothermic and therefore the interface of the $MAPbBr_3$ perovskite was energetically favorable to be passivated with K ions, which was well consistent with the experimental measurements reported recently (*23*); (2) The bond length of Pb-Br (where Br atom was the terminal one) was expanded from 3.095 Å of the bulk lattice to 3.31 Å of the mica/perovskite interface; (3) the distance of K-Br was about 3.25 Å, which was exactly equal to the ionic crystal KBr,(*24*) suggesting strong interactions between the potassium and the bromide at the mica/perovskite interface.

Structure relaxation of the mica/perovskite interface was confirmed through macroscopic XPS measurements. The $Pb_{4f}$, $Br_{3d}$ and $K_{2p}$ core levels were probed for two sheets with the 50nm and 7nm thicknesses, respectively. As shown in Fig. 2B and C, the Pb $4f_{7/2}$ core level peak located around 138.5 eV corresponded to the binding energy (B. E.) of Pb-Br bonds;(*25*) and the Br 3d peak at 68.6 eV corresponded to the B. E. of Br with Pb.(*26*) For the 7nm sheet, in comparison with the 50nm sheet, the B. E. of Pb and Br valence electrons were shifted to a higher energy, implying the structure relaxation of 2D $MAPbBr_3$ perovskites with the thickness reduction. We proposed two reasons attributed to the structure relaxation: the lattice expansion with thickness decreasing and the mica/perovskite interface relaxation. Thickness-induced lattice modification had been commonly observed in low-dimensional perovskites.(*27*) In the case of epitaxial 2D $MAPbBr_3$ perovskites on mica, with a high surface-to-volume ratio, the interface effect became more prominent, leading to $K_{2P}$ core level change in spectra. For the 7 nm perovskite, K $2p_{1/2}$ and $2p_{3/2}$ core level peaks located at 296.3 and 293.4 eV, respectively, which was consistent with the corresponding values of mica crystal,(*28*) indicating that most x-ray photoelectrons were derived from the mica substrate. With the thickness increasing, the photoelectrons from the substrate were substantially suppressed, thereby the XPS signals originated from the interface was enhanced. As a result, when thickness of the sample was increased, the $K_{2P}$ core level peaks would not only decrease in intensity but also shift to the lower energy, as clearly shown for the results obtained from the 50 nm perovskite. More importantly, the B. E. of K 2p for the 50 nm perovskite at 293.2 and 296.3 eV coincided exactly with the XPS fingerprint of K-Br bond in ionic crystal KBr.(*29*) This was a strong indication of direct K-Br interactions between the perovskite and mica substrate.

The thermodynamic stability of the mica/perovskite interface was also enhanced, which was supported by both DFT calculations and experimental observations. According to the classic epitaxy growth model, the lower interface energy $\gamma_{interface}$ was benefit for driving the 2D morphology, layer-by-layer growth of the films.(*30*) The calculated formation energy of the mica/perovskite interface was negative about -19.8~-43.2 meV/atom (Tab. S1), suggested the mica/perovskite interface was more stable than the bulk lattice. Therefore, it was reasonable to regard the K/Br interactions acted as the thermodynamic driving forces for 2D perovskite growths on mica. In growth experiments, we had observed that there was significant tendency of epitaxial growth of the $MAPbBr_3$ perovskite on the mica surface (video S2). The surface morphologies of the epitaxial perovskites with various thicknesses exhibited atomically flat (Fig. 3A~C, and Fig. S4~6 for $MAPbI_3$, $MAPbCl_3$ and $CsPbBr_3$ sheets). For the 8 nm and 35 nm 2D $MAPbBr_3$ perovskites, root means square roughness (rms) of surfaces were ~ 5 Å and ~4 Å, respectively, and their edges were sharp, which exhibited a typical layer-by-layer growth mode. When the thickness increased to 1 μm, the surface roughness of the perovskite became larger and rms was increasing to ~14 Å, suggesting the island nucleation appears. In addition, no steps and spiral-islands were founded in the perovskite surface suggesting the rare crystallographic defects in the epitaxial 2D perovskites. More discussion about the growth thermodynamic of 2D perovskites were available in the supplementary text (*20*).

We have confirmed that the K/Br interaction was the key point for the liquid epitaxy growth of 2D non-layered OIHPs, which determined their growth habits on mica were completely different from the cases on silicon, quartz and ITO glass. To further verify the generalizability of the K/Br mechanism, we attempted to grow 2D $MAPbBr_3$ perovskites on $KTiOPO_4$ (KTP) crystal. KTP crystal was a traditional ferroelectric.(*31*) The crystal structure was characterized by the $K^+$ channels, which were encircled by the helical chains of distorted $TiO_6$ octahedra and $PO_4$ tetrahedra.(*32*) These channels were along the *c*-aixs (001) direction, where $K^+$ ions could diffuse due to the vacancy. Whether 2D $MAPbBr_3$ perovskites could obtain the epitaxy growth on the $KTP_{(001)}$ substrate. In analogy to the growth on mica, $MAPbBr_3$ perovskites exhibited 2D growth on the $KTP_{(001)}$ substrate. However, their thicknesses were much larger as a result of coarse surface for KTP compared to the mica. The perovskites solidly attaching on KTP surface indicated the heteroepitaxy nature. Hence, it has validated that the K/B mechanism also worked in the case of the KTP substrate. (fig. S13).

**Optical Characteristics of 2D OIHPs**

To further demonstrate the applications in the opto-electronic area, we investigated the optical and photoelectrical properties of the epitaxial 2D $MAPbBr_3$ perovskites. As previously mentioned, the 2D $MAPbBr_3$ perovskites on mica had high-quality crystalline structure, such as atomically flat surface, uniform thickness, no spiral growth-steps, a vertical sharp-edge and so on, which were beneficial for yielding uniform PL emissions in plane, as shown in Fig. 3A-C. Owing to the quantum confinement and interface effects, the 2D $MAPbBr_3$ perovskites had unusual optical properties compared with their bulk counterpart. Both PL emission and optical absorption edge of 2D $MAPbBr_3$ exhibited blue shifts compared with the bulk crystals (Fig. 3D, E). The bulk $MAPbBr_3$ crystals had an emission peak at 545 nm (2.27 eV), and 2.14 eV bandgap estimated from optical absorption. For 2D $MAPbBr_3$ perovskites, PL peak was blue shifted to 519 nm and its bandgap increased to 2.44 eV. The DFT calculations had shown a 0.17 eV blueshift in bandgap for monolayered $MAPbBr_3$ perovskites, which was smaller than the blueshift of 0.30 eV obtained from experimental measurement.(*20*) This suggested that the enhancement of Coulomb interactions induced by the quantum confinement effect was also important for the 2D $MAPbBr_3$ perovskite with the 7 nm thickness.(*33*) Moreover, the exciton binding energy ($E_B$) of 2D $MAPbBr_3$ perovskite could be extracted from optical absorption,(*34*) which was 30 meV (Fig. 3E) larger than that of a bulk crystal (*35*), and also higher than the thermal energy (26 meV at 300K). Thus, the exciton was stable at room temperature and enhancing the luminescence emission of 2D $MAPbBr_3$ perovskites. Time-resolved PL spectrum indicated that the PL lifetimes of 2D $MAPbBr_3$ perovskites consisted of two components, the fast part of 6 ns and the low part of 50 ns (Fig. 3F) The former was attributed to the carrier recombination dominated by interface trap states, and the latter was governed by the recombination arising from their bulk phase. The optical spectrum deconvolution and exciton details are described in the Supplement text (*20*).

**Optoelectrical Characteristics of 2D OIHPs**

Figure 4A illustrates a photodetector fabricated with the epitaxial 2D $MAPbBr_3$ perovskite. Using a shadow mask pattern, 50nm Au metal was deposited as electrodes with a 100 µm gap. The thickness of the perovskite was about 20 nm, and the effective area of the device was about 50*100 $\mu m^2$. The 450 nm monochromatic light illuminating on the effective area just only be absorbed less than 5% by the 2D $MAPbBr_3$ perovskite. Figure 4b shows the light dependent I-V characteristics of the 2D $MAPbBr_3$ device. All *I-V* curves showed Ohmic conduction under illumination and exhibited extremely low dark current ~ $10^{-12}$ A. Under the bias of 5V and 10 V the photoresponsivity was measured, as shown in Fig. 4C. The photoresponsivity was increasing with the bias voltage increasing. A high photoresponsivity of 126 A/W was achieved. The carrier mobility of the 2D $MAPbBr_3$ perovskite was estimated to be 36.5 $cm^2V^{-1}S^{-1}$ using space-charge-limited current (SCLC) method (Fig. 4).(*36*) Using the oscilloscope, the temporal response of the 2D $MAPbBr_3$ photodetector was measured under a constant bias voltage of 5V with the indicate light intensity of 1.0 mW. The device had the on/off current ratio as high as $10^4$. Most importantly, achievement of fast response to the radiation had been demonstrated, with rise time and decay time of ~5 and 4.1 µs, respectively. The maximum of -3 dB bandwidth ($f_{-3dB}$) of the device could reach to 80 kHz, additionally, in experiments, the $f_{-3dB}$ value of certain device even exceeded the bandwidth of lock-in amplifier (0.1 MHz), which suggested the further device optimization would be improve performance of the 2D OIHPs device. The large optical absorption, small exciton binding energy, long carrier lifetime, combined with high-quality crystalline structure gave rise to excellent performances of 2D $MAPbBr_3$ perovskites. In addition, the trap states of the mica/perovskite interface (Supplement text, Fig. S13) were recognized to promote the photogain and carrier collection in a photodetector. Therefore, we could practically prove 2D $MAPbBr_3$ perovskites had the excellent photoelectronic performances better than that of the pristine monolayer transition-metal dichalcogenide photodetector.(*20*)

**Acknowledgments:** We thank X. M. Wu for his helpful discussion, L. Fu for the helps in the TRPL measurements, H. T. Zhou for providing the KTP substrate, and Y. L. Meng, L. Han and G. H. Hou for editing the supporting videos. **Funding:** This work was supported by the National Natural Science Foundation of China (51472123, 61871227). J. S. thanks the "Six Talent Peaks" Project of Jiangsu Province, China (No. R2016L07). Y. B. thanks the Research Innovation Program for College Graduates of Jiangsu Province, China (No. KYCX18_1031). **Author contributions:** Y. B., H. Z. and M. Z. performed the crystal growth, the photograph observation, the PL measurements, the TRPL measurements, the optoelectronic measurements and the AFM measurements. D. W. supervised the researches and analyzed the data. H. Z. contributed to the first-principles calculations. D. W. and H. Z. conceived the idea. H. X. and G. W. contributed to the instrumental developments. Y. X. and Y. Z. performed the optoelectronic measurements. H. J. performed the optical absorption measurements. R. P. and M. W. gave constructive discussions on the experiments. J. S. performed the sample preparations. D. W., H. Z., H. Y., Z. H. and Y. W. wrote the manuscript. **Competing interests:** The authors declare that they have no competing interests. **Data and materials availability:** All data needed to evaluate the conclusions in the paper are present in the paper and/or the Supplementary Materials. Additional data related to this paper may be requested from the authors.

**Supplementary Materials:**

Materials and Methods

Supplementary text

fig. S1. The $MAPbBr_3$ perovskites grown on the mica substrate without oleic acid.

fig. S2. The optical and fluorescent images of 2D $MAPbBr_3$ perovskites grown on the mica substrate with oleic acid.

fig. S3. Schematic illustration of the structure fracture of 2D $MAPbBr_3$ perovskites.

fig. S4. The liquid epitaxy of 2D $MAPbI_3$ perovskites on the mica.

fig. S5. The liquid epitaxy of 2D $MAPbCl_3$ perovskites on the mica.

fig. S6. The liquid epitaxy of 2D $CsPbBr_3$ perovskites on the mica.

fig. S7. Rocking-curve spectra for the (002) face of the 2D $MAPbBr_3$ perovskite on mica.

fig. S8. XPS spectrum of the 2D $MAPbBr_3$ perovskite on mica.

fig. S9. Perovskite grow on the ITO, $SiO_2$/Si and quartz substrate.

fig. S10. Schematic illustration of the $CH_3NH_3^-$ molecular rotation along with the (010) axis.

fig. S11. The structures of the K ions /perovskite interfaces with various spatial orientation of the $CH_3NH_3^-$ molecular.

fig. S12. Polar coordinates of totally interface energy vs the $CH_3NH_3^-$ orientation.

fig. S13. The liquid epitaxy of the 2D $MAPbBr_3$ on the KTP substrate.

fig. S14. The calculated band structures of the bulk $MAPbBr_3$ perovskite, 2D $MAPbBr_3$ perovskite and the perovskite surfaces.

fig. S15. The carrier recombinations in 2D $MAPbBr_3$ perovskites.

tables S1. The formation energy of the K/perovskite interface calculated by DFT.

tables S2. Device performance comparisons of the liquid epitaxy 2D $MAPbBr_3$ perovskite photodetector in this work with the reported perovskite-based photodetectors.

movie S1. The $MAPbBr_3$ sheet growth on quartz substrate.

movie S2. The liquid epitaxy of 2D $MAPbBr_3$ perovskite on mica.

References (*37-48*)

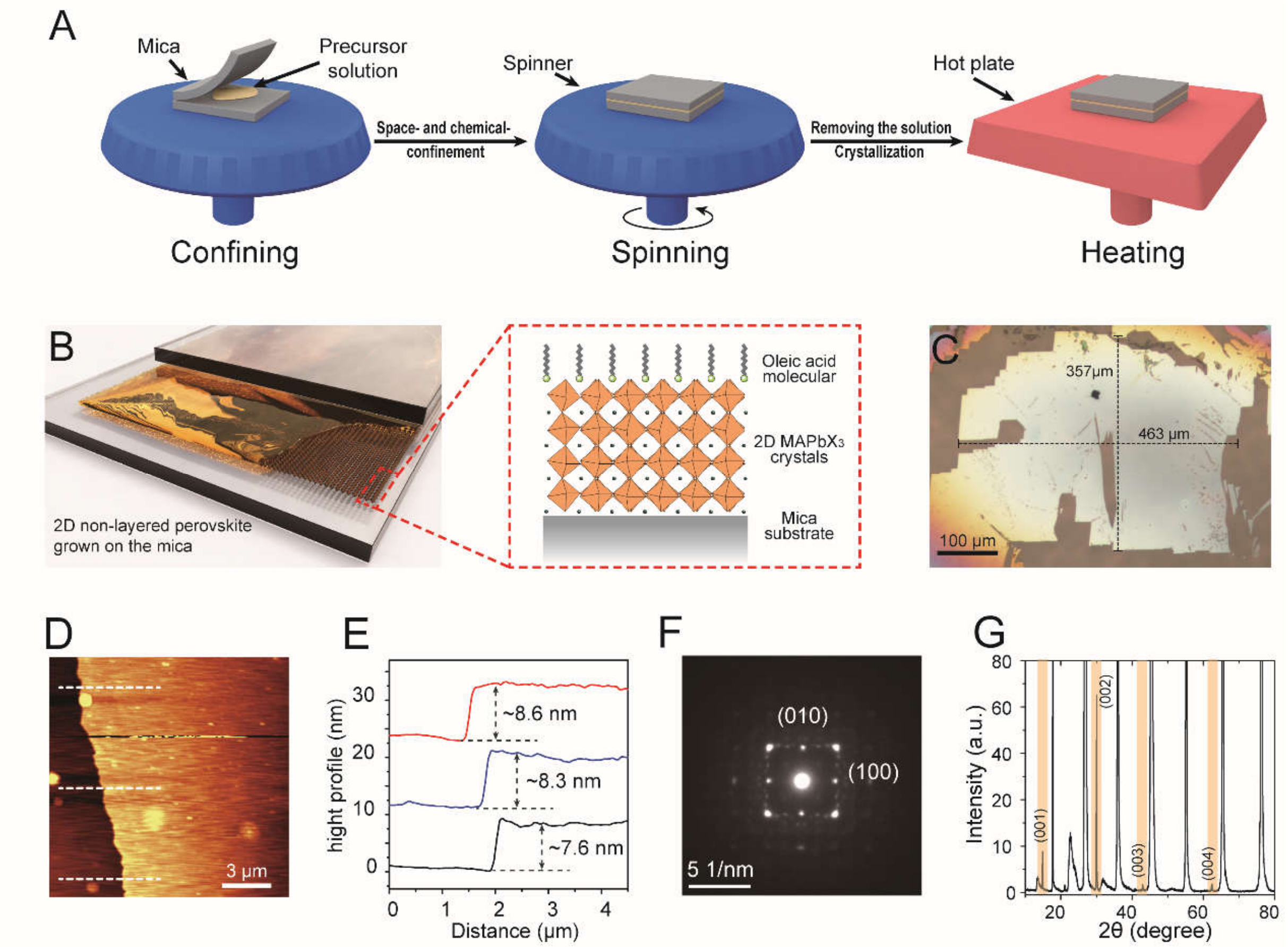

**Fig. 1. Liquid-phase epitaxy growth, morphological and phase characterization of 2D non-layered $MAPbBr_3$ perovskites on mica substrate.** (**A**) Schematic illustration of the growth procedure of 2D non-layered perovskite on mica substrate. (**B**) Schematic of 2D perovskites grown between two freshly-cleaved micas, and the interface structure of the as-grown perovskites. (**C** and **D**) Optical microscopy and AFM images of 2D $MAPbBr_3$ perovskite with the size of several micrometers and the thickness of 8 nm. (**E**) The height profile of perovskites along the white dash line in (D). (**F** and **G**) The electron diffraction pattern and grazing incidence XRD of the epitaxial 2D $MAPbBr_3$ perovskites, indicating their anisotropy crystal growth, i.e. the in-plane growth along with the (100) and (010) orientation but the confinement in the (001)

orientation. On the other hand, XRD data suggests the high crystalline quality of 2D $MAPbBr_3$ perovskite.

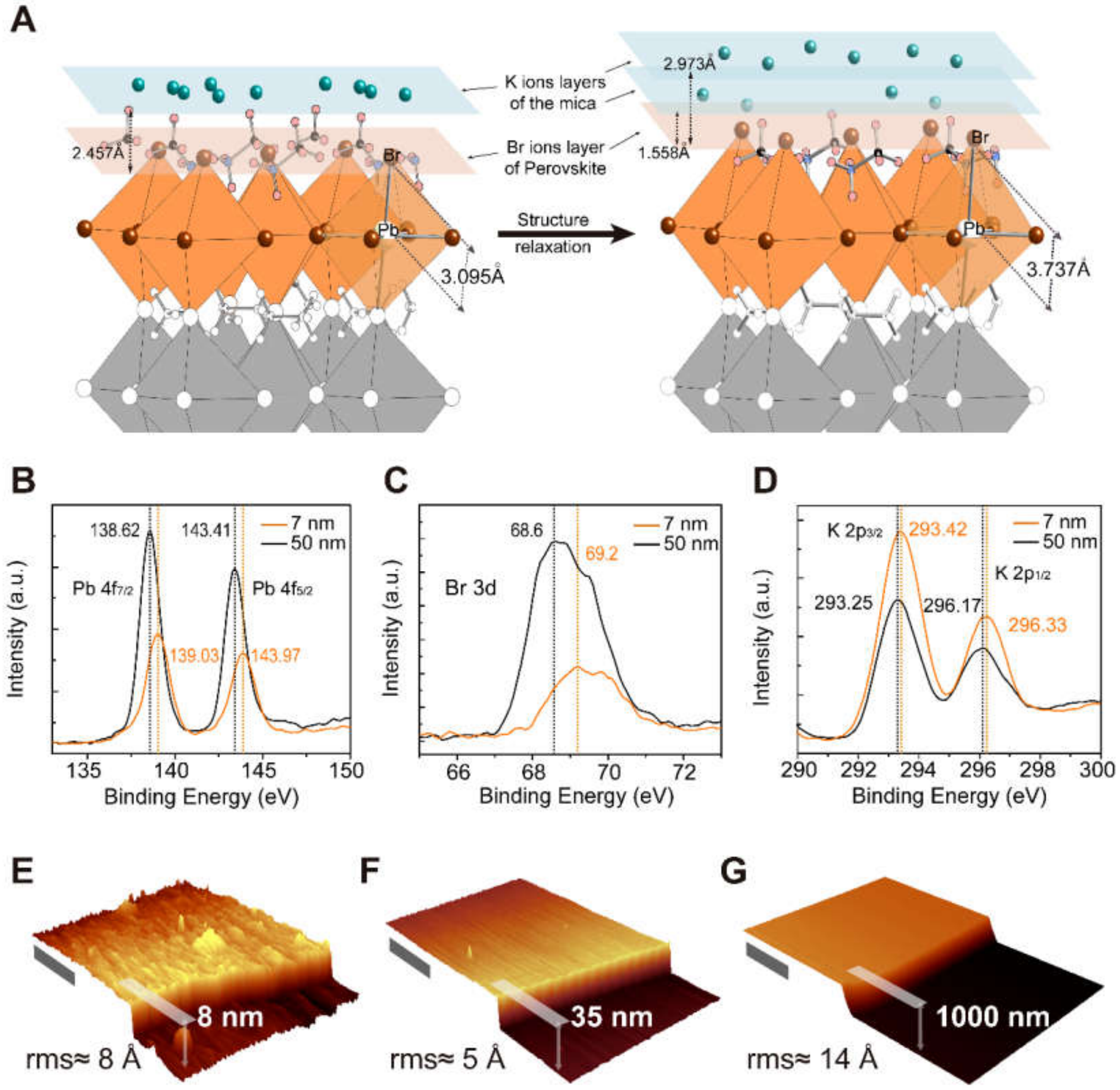

**Fig. 2. The structure relaxation of the mica/perovskite interface, and the surface morphologies of the epitaxial 2D perovskites.** (**A**) Schematic illustration of the structure relaxations of the perovskite/mica interface before and after the K ion absorption. (**B** to **D**) XPS spectra of Pb 4f, Br 3d and K 2p for two 2D perovskites with the thickness of 7 nm and 50 nm, respectively. (**E** to **G**) 3D AFM images exhibit the surface morphologies and high profiles of

three $MAPbBr_3$ perovskite sheets with the thicknesses of 8 nm, 35 nm and 1000 nm. The rms of three sheets is ~8 Å, 5 Å and 14 Å, respectively. Scale bar, 7 μm.

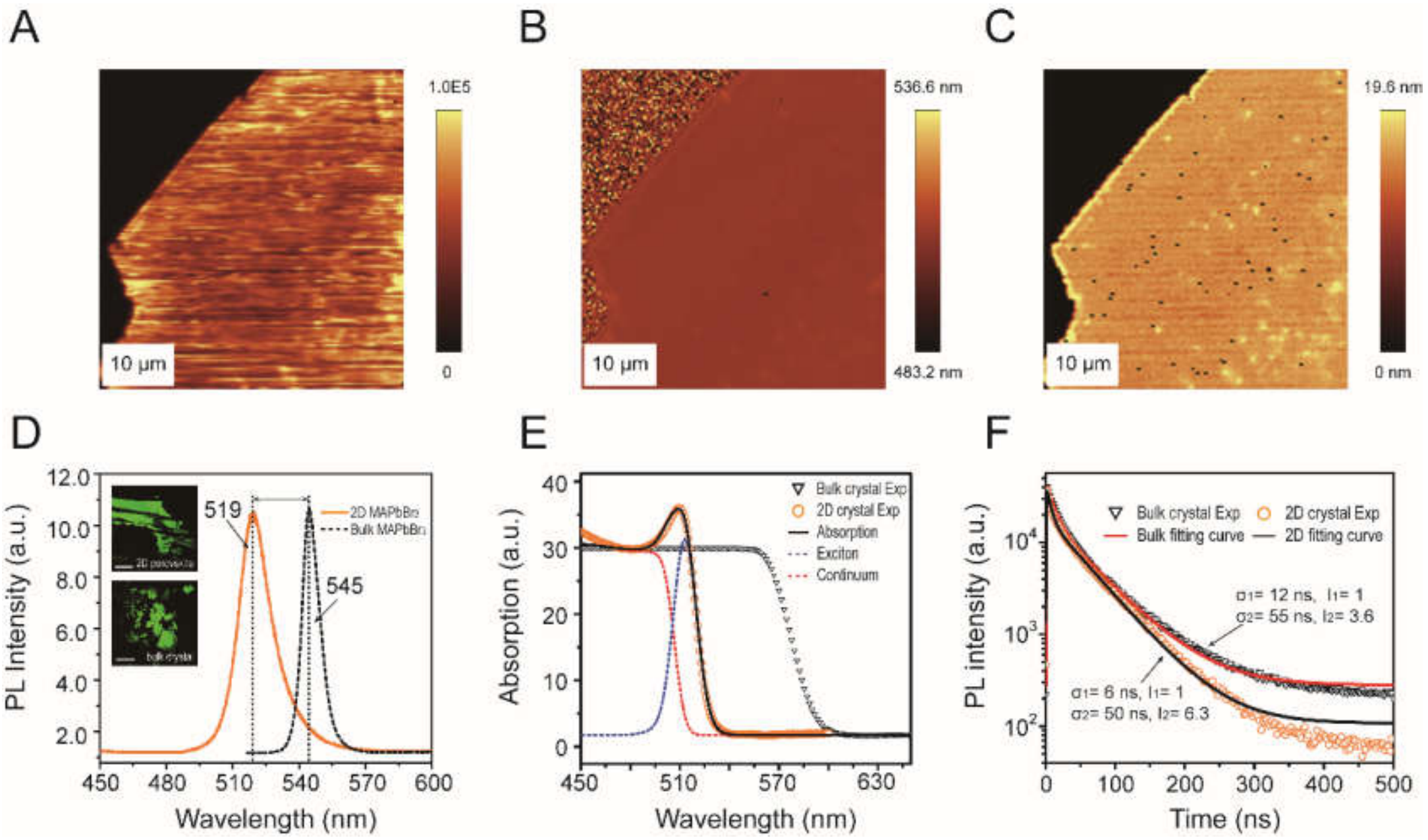

**Fig. 3. Optical characteristics of the epitaxial 2D perovskites.** (**A** to **C**) The intensity, peak wavelength and FWHM of the PL emission for the 2D $MAPbBr_3$ perovskite on the mica substrate, indicating its uniform illumination in plane. (**D**) PL curve of the 2D $MAPbBr_3$ perovskite thinner than 10 nm (the orange solid line), blue-shifted by 35 nm compared with the bulk crystal (the black dash line). Insets of (D) are the fluorescence image of the 2D and bulk perovskite, respectively. (**E**) Optical Absorption of the 2D $MAPbBr_3$ perovskite compared with the bulk crystal. The optical absorption of the 2D perovskite is deconvoluted according to the Elliott's theory of Wannier excitons. The blue dash line: exciton transitions. The red dash line:

band-to-band transitions. (**F**) The decay times of the PL emissions at 519 nm for the 2D perovskite and 545 nm for the bulk crystal, respectively.

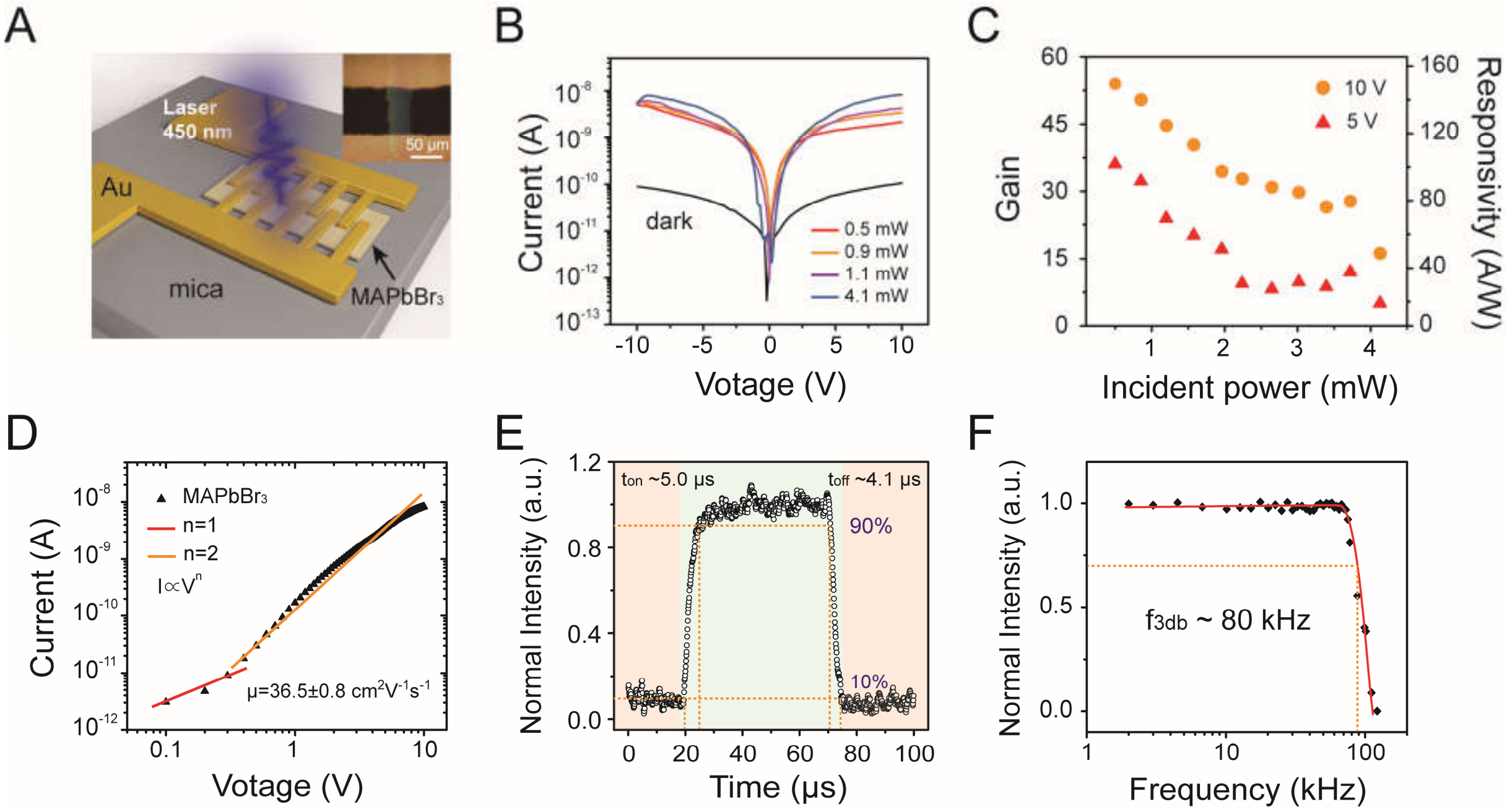

**Fig. 4. Photoelectronic performances of the epitaxial 2D $MAPbBr_3$ perovskite.** (**A**) Scheme of the 2D perovskite-based photodetector. Inset of (A) is the optical microscopy image of the device. (**B**) I-V curves of the 2D perovskite-based photodetector measured in the dark and under various light illumination, respectively. (**C**) Gain and photoresponsivity of the device at 5V and 10 bias, respectively. (**D**) The I-V curve of the device at 5V bias under 0.9 mW illumination. The carrier mobility in 2D perovskite is estimated to be 36.5±0.8 $cm^2V^{-1}S^{-1}$ using the SCLC method. (**E** and **F**) Time-resolved response and 3 dB bandwidth of the device at 5V bias.

# Supplementary Materials for

## Liquid Epitaxial Growth of Two-dimensional Non-layer structured hybrid Perovskite

Yu Bai, Haixin Zhang, Mingjing Zhang, Di Wang*, Hui Zeng*, Hao Xue, Guozheng Wu, Ying Xie, Yuxia Zhang, Hao Jing, Jing Su, Haohai Yu, Zhanggui Hu, Ruwen Peng, Mu Wang, Yicheng Wu

Correspondence to: diwang@tjut.edu.cn (D. W.); zenghui@njust.edu.cn (H. Z.).

**This PDF file includes:**

Materials and Methods
Supplementary Text
Figs. S1 to S15
Tables S1 to S2
Captions for Movies S1 to S2

**Other Supplementary Materials for this manuscript include the following:**

Movies S1 to S2

## Materials and Methods

### MM1. The liquid epitaxy and device fabrication of 2D non-layer structured perovskite

The single-crystalline growth solution was prepared using the crystallization techniques described in our previous works (*37*), which was demonstrated to be an effective month solution to grow high-quality hybrid organic-inorganic perovskite (HOIP) crystals. For the $CH_3NH_3PbBr_3$ preparation, the first step was the methylamine bromide synthesis by reacting methylamine, 40 wt% in aqueous solution, with bromic acid (HBr), 48 wt% in water. Typically, a mixture of 7 ml HBr and 9 ml methylamine were taken in to a round bottom flask and stirred at 0 °C for 2 h, and then the solvent was removed using the rotary evaporator until a white MABr powder crystallized. The obtained salt was washed and filtered three times with diethyl ether. The second step, the prepared MABr powder mixed with $PbBr_2$ at 1:1 mol ratio in N, N-dimethylformamide (DMF). The mixed solution was stirred for 12 h until the reactions dissolved completed, and then added oleic acid (OA) with different amounts. For $CH_3NH_3PbCl_3$ and $CH_3NH_3PbI_3$ crystalline, the solvents were a dimethylsulfoxide (DMSO) and γ-butyrolactone (GBL), respectively. All the reactions were purchased from Alfa-aesar.

Ultrathin 2D perovskite sheets were epitaxially grown on the fresh cleaved mica substrates using space- and chemical- confinement method. First, the mica was cleaved into two pieces using a surgical blade. One piece was used as the substrate, the other was used as the cover to confine the vertical growth of perovskite. The prepared $MAPbBr_3$ solution (dissolved in DMF at a concentration of 0.2 M) was dipped on the fresh cleaved surface of the mica substrate, and then the other mica was covered onto this substrate. The combination was fastened using a tape, and span at 3000 r. p. m. for 30s in order to reduce the solution thickness. Then, the combination was placed on the heating stage for 2h, and ultrathin 2D perovskite sheets with the thickness at the nanometer scale and the size of several micrometer could be obtained on the mica substrate. The OA in solution could attach the surface of the grown perovskite, which confine the crystal growth at the molecular level. Finally, a 50-nm-thick gold electrode was deposited by thermally evaporation with a shadow mask. The completed devices were stored in a Nitrogen environment for further use.

### MM2. Structure and morphology

The bright field and fluorescence images of the 2D hybrid perovskite sheets were observed with Olympus optical microscope (BX53-fluorescenc). The fluorescence mode of BX53 uses a fluorescence filter with a medium sized band-pass excitation filter (450-490nm) and a long-pass barrier filter (515nm cut-on). The thickness and surface morphology of the 2D perovskite sheets were measured using atom force microscopy (WITec, alpha 300 A for Atomic Force Microscopy (AFM)) in tapping mode. The as-grown 2D hybrid perovskite were identified using powder x-ray diffractometer (Rigaku MAX-RD). The crystal lattice was analyzed the electron diffraction pattern, which obtained by transmission electron microscopy (TEM) (JEOL, JEM-1011). TEM samples are prepared using lacey carbon Cu grid to scratch the surface of the mica substrate. X-ray photoelectron spectroscopy (XPS) was performed on a Thermo Scientific ESCALAB250Xi, with a mono-chromated Al X-ray source. The power was 150 W and spot size was 50 micrometres.

### MM3. Optical characterization

Steady-state photoluminescence (PL) spectra were measured with a confocal micro-Raman system (WITec, alpha 300 A). The sample was excited with a fiber-coupled laser source with the 405 nm wavelength. The laser beam was focused by a microscope into a small focal volume on the surface of the sheets. The typical spot size was about 2 μm. The laser power kept less than 1mW to avoid laser-induced damage on the sample surface. Time-resolved PL spectra were carried out by time-resolved confocal fluorescence microscopy (PicoQuant, MicroTime 200). The Steady-state optical absorption is carried out with a micro-spectrophotometer (CRAIC, 20/30 PV Tm). To eliminate the influence of mica substrate, we calibrate the result by measuring the absorption spectra of the same mica substrate. All experimental measurements are carried out at room temperature.

**MM4. Device characterization**

The 2D perovskite-based photodetector was characterized using a home-built Photocurrent Analyzer. The monochromatic light at 450 nm wavelength was provided by a semiconductor diode laser (Thorlabs, LP450-SF15), and the light intensity was calibrated using an optical power meter (Thorlabs, S121C). The photocurrent of the detector was first amplified by a low noise current preamplifier (Stanford, SR570), and then the output signal is recorded using a digital source meter (Keithley, 2612B). The time dependent photoresponse signal was recorded using a digital oscilloscope (National Instruments, PXI-5122) with 50 ohm input impedance. The 3dB bandwidth was measured using a lock-in-amplifier (Stanford, SR830). All measures were performed using a probe station on the floating table to minimize vibrational noise.

**MM5. The density functional theory (DFT) calculation**

The electronic structures of the perovskite crystal and the perovskite/mica interface were carried out using Vienna *ab initio* simulation package (VASP). The projector augmented wave (PAW) pseudopotentials were used for the electronic exchange correlation potential. The cutoff energy of the wave functions was set to 400 eV. The exchange-correlation potential was described by the generalized gradient approximation (GGA). The vdW-DF dispersion correction was introduced to describe the van der Waals interaction of the organic molecular. The convergence criteria for the ionic relaxation was set in $10^{-4}$ eV per atom and the remnant force of each atom was smaller than $3\times10^{-3}$/Å. For the estimation of the perovskite/mica interface, the perovskite (001) slab model contains four layers, and a vacuum distance was set to be 25 Å to eliminate the interactions of the adjacent layer. Owing to the large lattice mismatch, it was difficult to build the interface model of the perovskite and mica. In this work, the equilibrium K atoms (about 1.5 K atoms for per unit cell of the perovskite) was placed on the perovskite (001) surface to serve as a model of the mica (001) surface. Consequently, we have performed the calculations with various K adatom adsorbed on the perovskite surface. The K adatom and the top Pb-Br octahedra layer were allowed to relax, while the three bottom layers were fixed. The Monkhorst-Pack set of 9×9×1 k-grid is used to sample the Brillouin zone (BZ) for the geometry optimization, and a denser sampling of 15×15×1 is used for electronic structure calculations.

To investigate the orientation-dependent free energy of the perovskite/mica interface, the MA molecular in the perovskite were rotated 360 degree along with b axis (fig. S9). The rotation step was set to 30 degree, and the free energy and structure of the interface were calculated.

**Supplementary Text**

**ST1.** The liquid epitaxy for 2D non-layer structured perovskite.

Mica is an excellent substrate for van der Waals epitaxy growth of 2D materials. Our experimental observations suggested the interactions between perovskite and mica were not the van der Waals forces but the strong chemical bond of K/halogen. At first, 2D perovskites on mica did not exhibit the oriented growth, as justifying by the growth monitor movie. Mica is a typical layered structure crystal, in which the layers connect with each other via the weak Van der Waals forces. Therefore, mica is considered to be an idea substrate to grown 2D materials, such as $MoO_3$, $MoS_2$. The epitaxial materials on mica usually exhibit the oriented growth along with three directions (60 degree to each other). However, in our experiments, we noticed that 2D perovskites randomly distributed on the mica surface without any favorable orientations (fig. S2). Our DFT calculations suggested that the stabilities of the perovskite/mica interface were almost unaltered for various perovskite orientations due to the rotation of the $CH_3NH_3^-$ molecular (fig. S10-S12). Second, the epitaxial perovskites were strongly stuck to the mica substrate. During the van de Waals epitaxy, the forces between 2D materials and mica is intuitively considered to be the van de Waals force, thereby the 2D materials are easily exfoliated from the mica substrate. In the case of the perovskites on mica, we experimentally found that 2D perovskites were strongly stuck to the mica surface. When removing the mica cover of the growth system, some grown 2D perovskites were teared and then strongly adhered on the top and bottom mica, respectively (fig. S3). The DFT calculations had revealed the observed strong adhesion force derived from the chemical bond between K/halogen interaction connecting the mica and perovskites, which promoted the epitaxy growth of perovskites on the mica and KTP crystal.

ST2. Optical absorption deconvolution and exciton binding energy of 2D $MAPbX_3$ perovskite.

Using the Elliott model, the band bandgap $E_g$ and exciton binding energy $E_B$ can be accurately estimated basing on the optical absorption. The model suggests the absorption coefficient $\alpha(\hbar\omega)$ is associated with continuum absorption and exciton absorption involving the bandgap $E_g$ and the exciton binding energy $E_B$ as:

$$\alpha(\hbar\omega) \propto \frac{\mu_{CV}^2}{\hbar\omega}\sqrt{E_B}\left[\sum_n \frac{4\pi E_B}{n^3}\delta(\hbar\omega - E_n^B) + \frac{2\pi\sqrt{E_B}\,\theta(\hbar\omega - \mathrm{E_g})}{1 - e^{-2\pi\sqrt{\frac{E_B}{\hbar\omega - E_g}}}}\right] \quad (1)$$

where $\mu^2_{CV}$ is the squared transition dipolemoment, $E_g$ is the bandgap, $E_B$ is the exciton binding energy, $E_n^B$ is the energy of nth state. The first term in eq. (1), the delta function $\delta(\hbar\omega - E_n^B)$, represents the excitonic transition to the nth state with energy $E_n^B = E_g - \frac{E_B}{n^2}$. Because excitons cannot be spectrally resolved for n > 1 at room temperature, in this work, all fitting applies for n =1 exciton transition. The second term is a Heaviside step function $\theta(\hbar\omega - \mathrm{E_g})$, which represents the continuum band-to-band transition above bandgap $E_g$. To describe the line broadening and the non-parabolic dispersion of energy band in experiments, a Gaussian function is convoluted with the delta function and the Heaviside step function, respectively.

In this work, the $E_B$ values estimated from optical absorption were 30 meV and 10 meV for 2D $MAPbBr_3$ and $MAPbI_3$ perovskites, respectively. These values were much smaller than that of the layered perovskite, which was evaluated to be larger than 100 meV. The tiny $E_B$ was attributed to the intrinsic non-layer structure of 2D $MAPbX_3$ perovskites. The inorganic components have larger high frequency dielectric constant compared with the layered organic molecules, which minimize the dielectric confinement effects. According to Bohr model, the exciton energy could be expressed as $13.6eV\frac{\mu}{m_0\varepsilon_r^2}$, where $m_0$ was the electron static mass, the dielectric constant of $\varepsilon_r$ was about 25 and the electron effective mass μ was about $0.2m_0$ for $MAPbBr_3$. It gave rise to the exciton energy of about 4.3 meV in theory, which was smaller than the experimental measurement of 30 meV. The distinct differences were attributed to the increasement of the dielectric constant for 2D $MAPbBr_3$ perovskites ($\varepsilon_r$~30 estimated from the binding energy in this work). Hence, the exciton Bohr radius of 2D $MAPbBr_3$ perovskites was theoretically anticipated to be 7.9 nm, comparable to the thicknesses of as-grown $MAPbBr_3$ perovskites in experiments. For 2D $MAPbI_3$ perovskites, the effective mass μ was about $0.093m_0$. According to the experimental $E_B$ of 10 meV, Bohr radius of 2D $MAPbI_3$ perovskites was estimated to be about 3.0 nm. In this work, the measured 2D $MAPbI_3$ perovskites had the thinnest thickness of 6 nm, which was thicker than its Bohr radius. Eventually, the exciton binding energies of both 2D $MAPbBr_3$ and $MAPbI_3$ perovskite were essentially on the order of $K_BT$ at room temperature. So, excitons in our perovskite samples were easy to be separated at room temperature, which might be responsible for the high performance of 2D $MAPbBr_3$ photodetector.

ST3. Discussion on the optical properties of 2D $MAPbBr_3$ perovskite.

The spectroscopy measurements showed a large blue shift of the PL emission for 2D $MAPbBr_3$ perovskite, as compared to their bulk counterparts. We discussed the possible reasons to the blue shift of PL emission. One reason was the band gap increasing with respect to the thickness reduction due to quantum confinements. To confirm this, the electronic band structures of the bulk and monolayer perovskites had been calculated. The calculated $E_g$ for the bulk $MAPbBr_3$ perovskite was 2.14 eV, which was in good agreement with the experimental observation (Fig. 3E). For the monolayer $MAPbBr_3$ perovskite, the calculated $E_g$ was increased to 2.29 eV (fig. S14). As a result, PL emissions of 2D $MAPbBr_3$ perovskite would shift to the high-energy side. However, the calculated 2.29 eV of the monolayer $MAPbBr_3$ perovskite was smaller than the experimental 2.44 eV of 2D perovskite (Fig. 3). This distinction suggested that other factors also had significant influence on the photoluminescence process, such as the quench of Coulomb screening and the enhancement of exciton binding energies induced by quantum confinement of the 2D nanostructure, or the traps states on the interface.

In ST2, the Bohr radius of $MAPbB_3$ and $MAPbI_3$ were estimated to be 7.9 and 3.0 nm, respectively. 2D perovskites as thin as Bohr radius should exhibit strong quantum confinement effect, such as blue shift of PL emission, large exciton binding energy. Here, we discussed other possible reasons to affect PL emissions of 2D perovskites, namely, the trap states. The DFT calculations had indicated that the mica/perovskite interface would introduce trap states into the energy band of perovskites. The electronic band structures of the perovskite/mica interface with various K adatoms were calculated and shown in fig. S13. We noted that the electron trap states mainly contributed by K adatoms were observed in the conduction band. In addition, the trap states located at Γ point of the K-space, whose energies were larger than the minimum of the conduction band. The electrons released from the traps and then transited to the band edge with the help of phonons. The top of the valence band was also at Γ point, but the bottom of the conduction band located nearby the Γ point, whose energy was about 0.078 eV lower than the conduction band edge of the Γ point (fig. S15). As a consequence, the carrier recombination at the band edge would emit shorter wavelength than that of the bottom band. Our DFT calculations had shown that the trap states at the interface were anticipated to induce blue shift of PL emission. In addition, owing to the momentum conservation at Γ point, electrons tended to fall back to the valence band by radiative recombination. The fast component of PL lifetime thus could be attributed to the carrier recombination at Γ point. 2D $MAPbBr_3$ perovskites had a high surface-to volume ratio, thereby the trap states strongly influenced the carrier recombination. The PL emission at 519 nm of 2D $MAPbBr_3$ perovskites was 0.11 eV larger than 545 nm of their bulk counterpart, which was consistent with our DFT calculations.

**Fig. S1.**

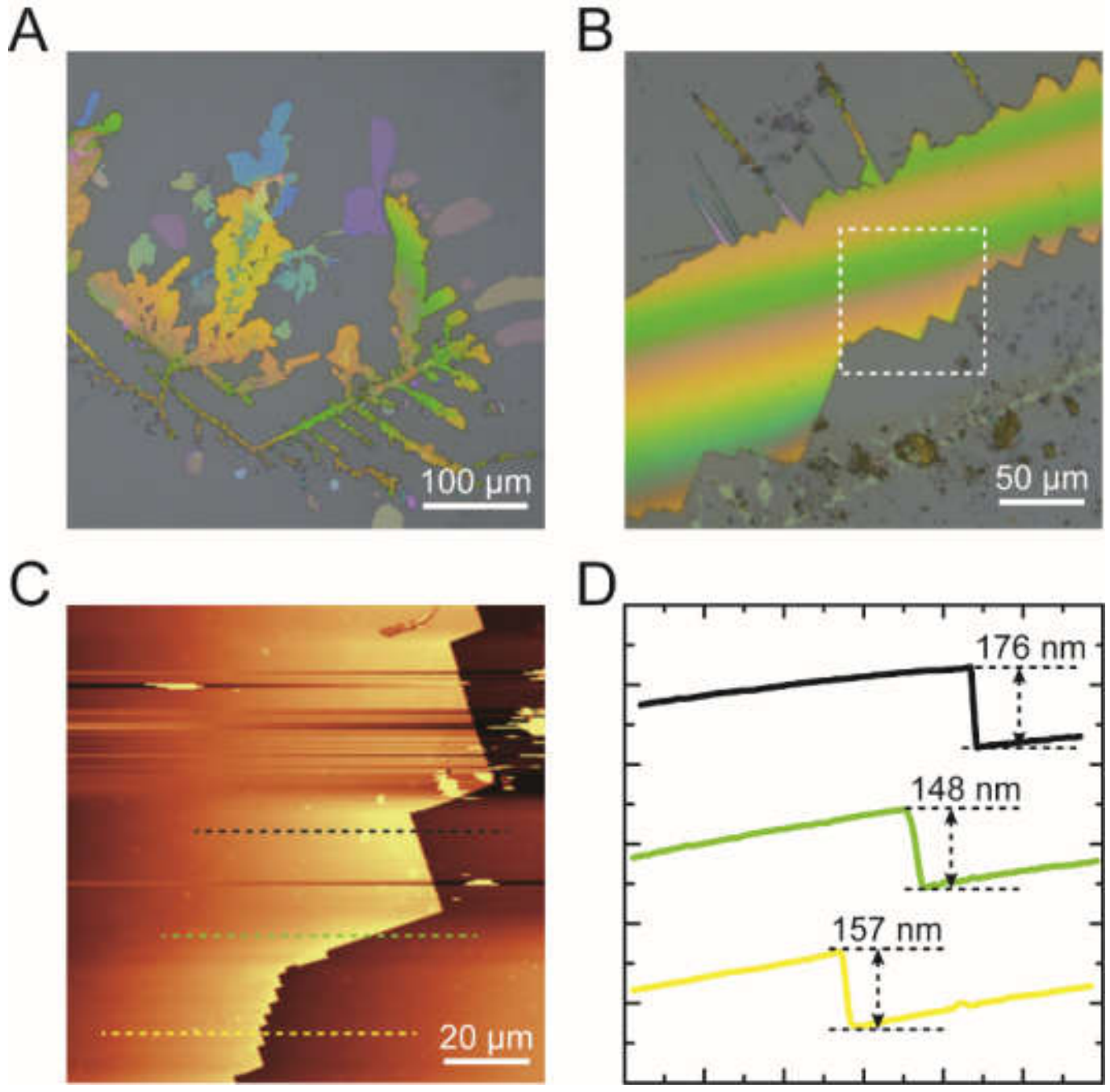

**Fig. S1. The $MAPbBr_3$ perovskites grown on the mica substrate without oleic acid.** (**A, B**) Optical images. (**C**) AFM image of the perovskite denoted by the dotted box in (B). (**D**) High profile denoted by the black, green and yellow dash lines in (C). We note that the $MAPbBr_3$ perovskite sheets grown without oleic acid are as large as several hundreds, but it is hard to obtain the sheets thinner than 100 nm thick.

Fig. S2.

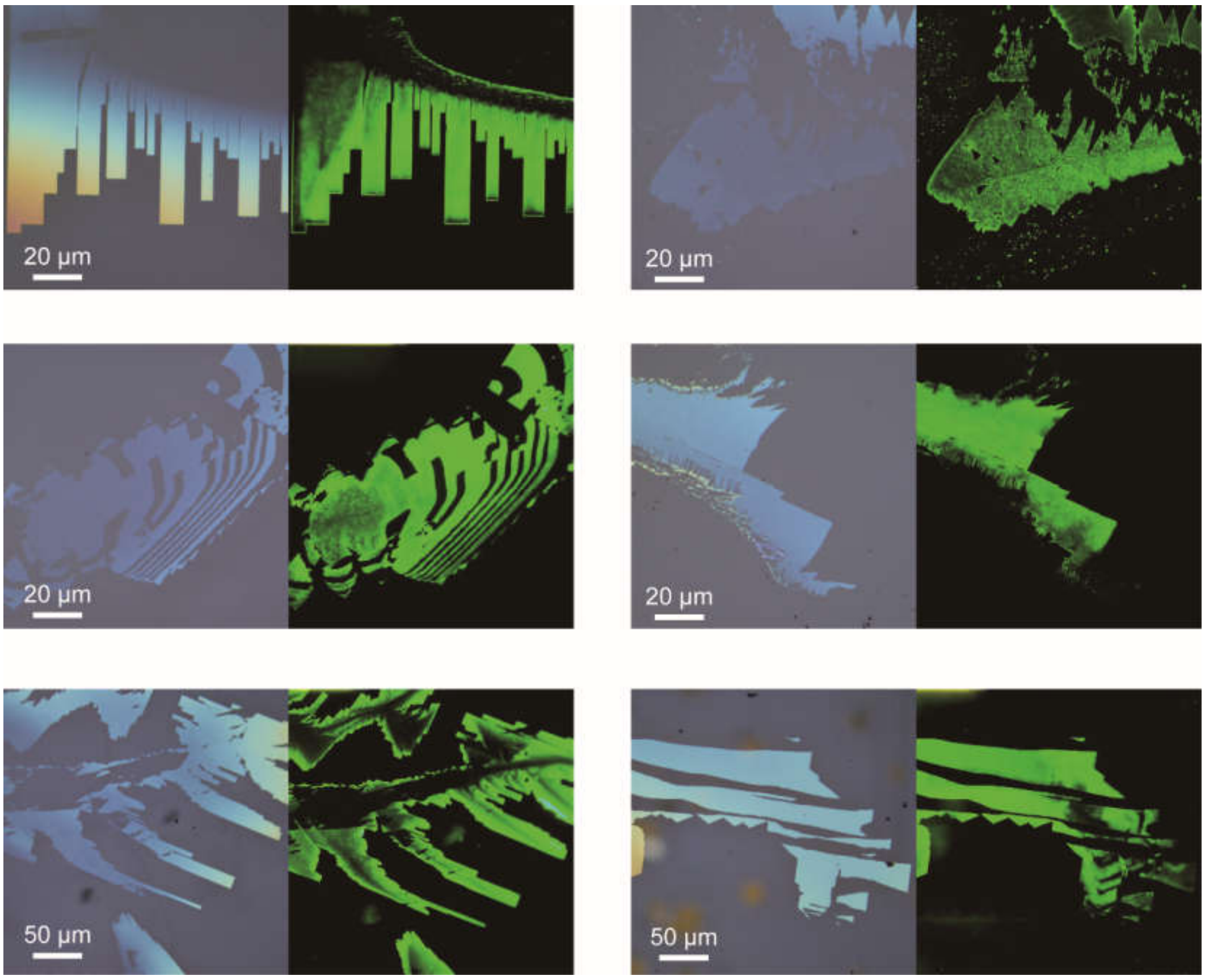

**Fig. S2. The optical and fluorescent images of 2D $MAPbBr_3$ perovskites grown on the mica substrate with oleic acid.**

Fig. S3.

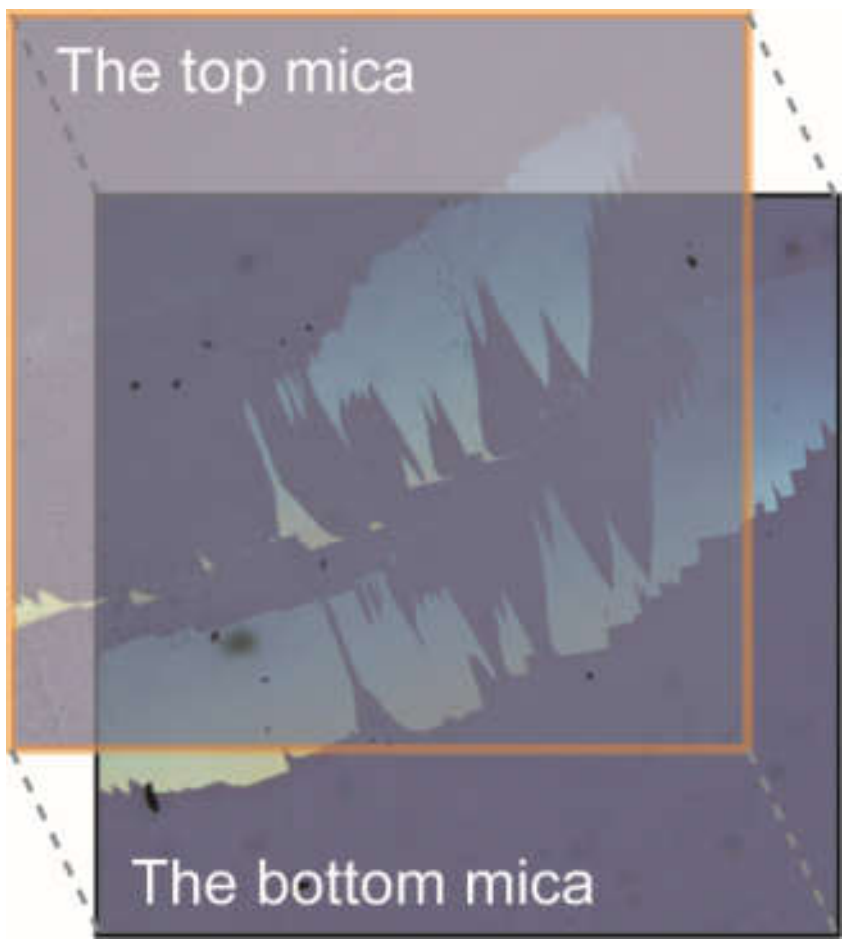

**Fig. S3. Schematic illustration of the structure fracture of 2D $MAPbBr_3$ perovskites.** The perovskite growth is confined by two mica substrates, and then it can epitaxy grow on the surfaces of top and bottom micas. When the top mica is removed from the bottom one, some parts of the grown perovskite will adhere to the top mica and make the film fracture.

**Fig. S4.**

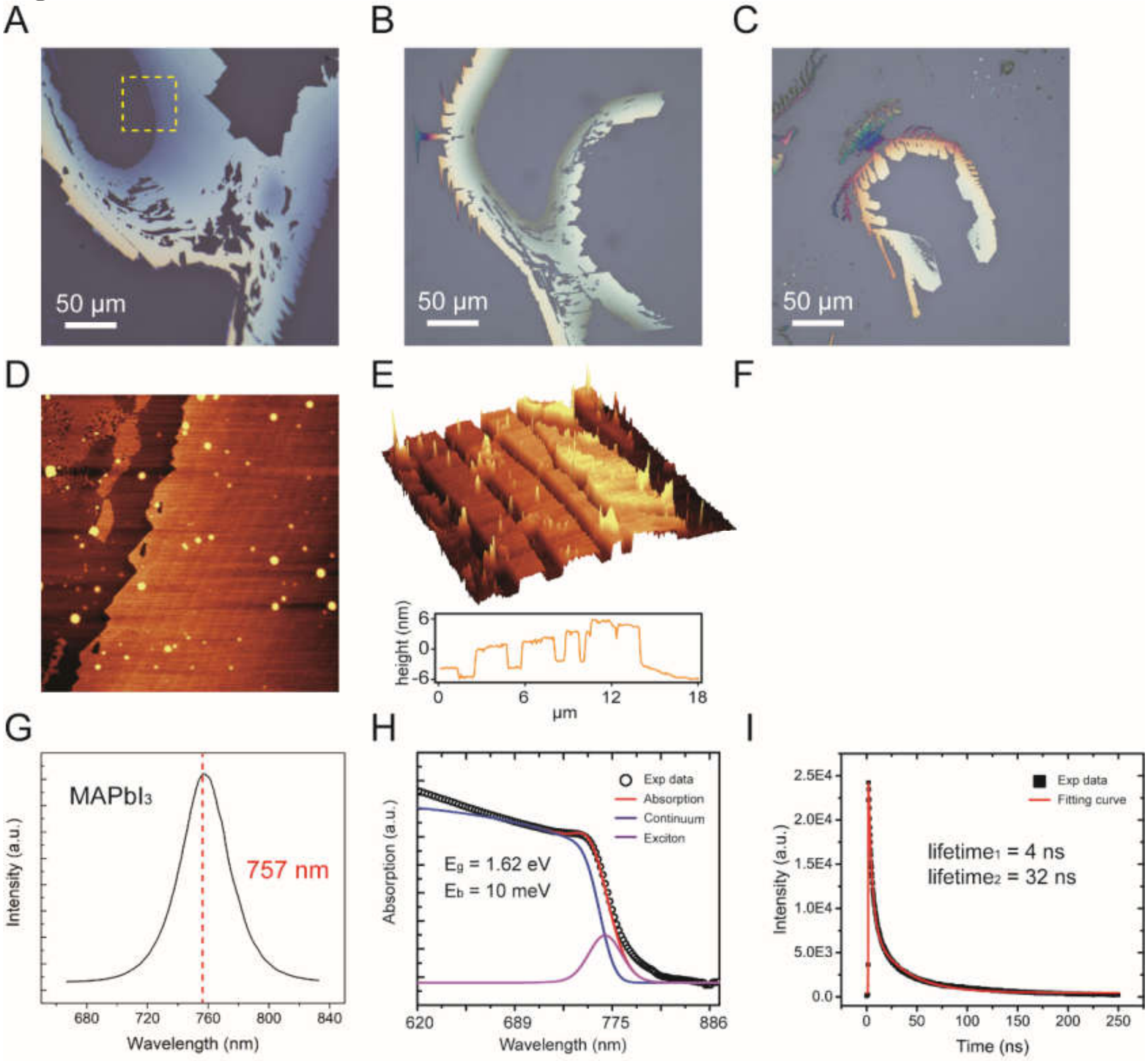

**Fig. S4. The liquid epitaxy of 2D $MAPbI_3$ perovskites on the mica.** (**A, B** and **C**) Optical images. (**D**) AFM image of the MAPbI3 perovskites denoted by the yellow dashed box in (A). (**E**) 3D image and high profile of the grown 2D $MAPbI_3$ perovskites. The grown 2D sheet with the thickness of 6 nm, and have the very sharp edge and smoothing surface. (**G**) PL of the 2D $MAPbI_3$ perovskites, which has a blue shift to the bulk $MAPbI_3$ perovskites. (**H**) Absorption of the 2D $MAPbI_3$ perovskites. According the Eillot model, the bandgap $E_g$ and the exciton binding energy $E_b$ can be estimated to be 1.62 eV and 10 meV, respectively. (**I**) The lifetime of the 757 nm peak, which is estimated to be 4 ns and 32 ns.

**Fig. S5.**

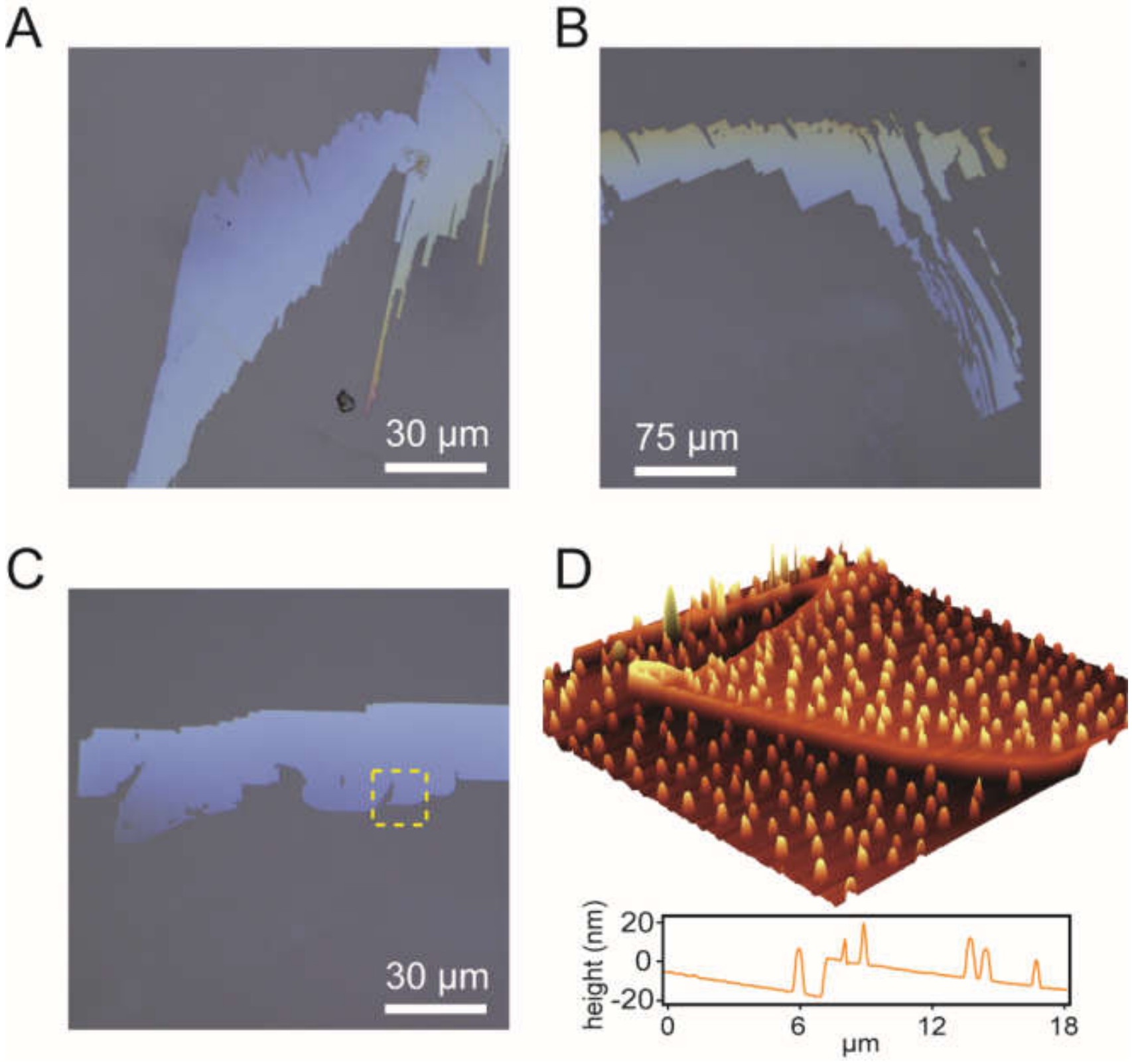

**Fig. S5. The liquid epitaxy of 2D $MAPbCl_3$ perovskites on the mica.** (**A, B** and **C**) Optical images. (**D**) 3D AFM image and the high profile of the 2D $MAPbCl_3$ perovskite denoted by the yellow dashed box in (C).

**Fig. S6.**

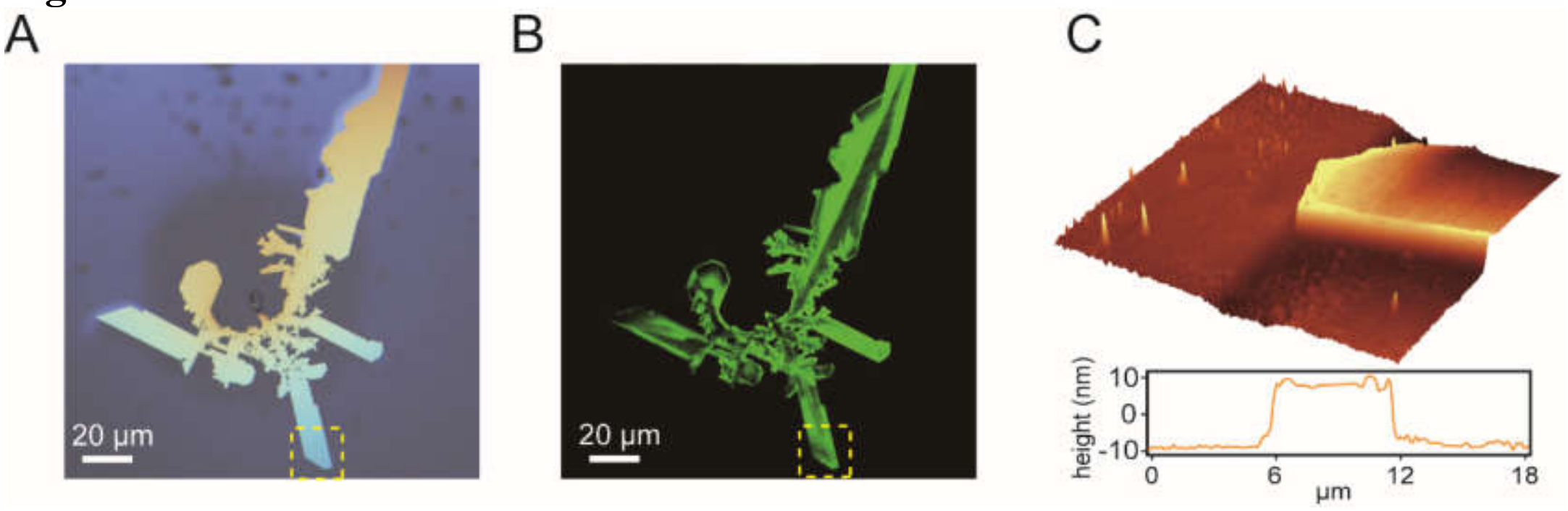

Fig. S6. **The liquid epitaxy of 2D $CsPbBr_3$ perovskites on the mica.** (**A, B**) Optical and fluorescent images. (**D**) 3D AFM image and the high profile of the 2D $CsPbBr_3$ perovskite denoted by the yellow dashed box in (A, B).

**Fig. S7.**

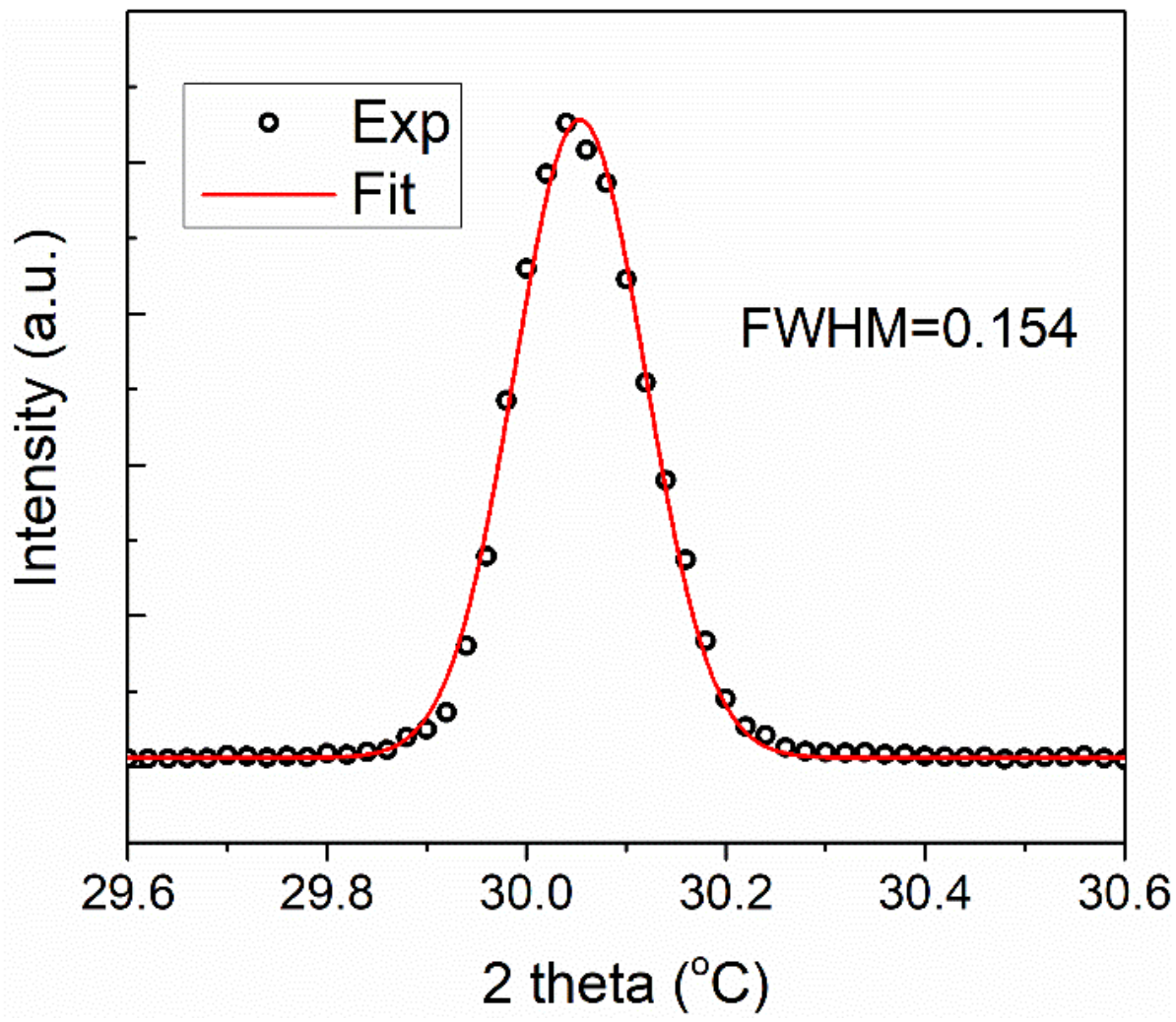

**Fig. S7. Rocking-curve spectrum for the (002) face of the 2D $MAPbBr_3$ perovskite on mica.**

**Fig. S8.**

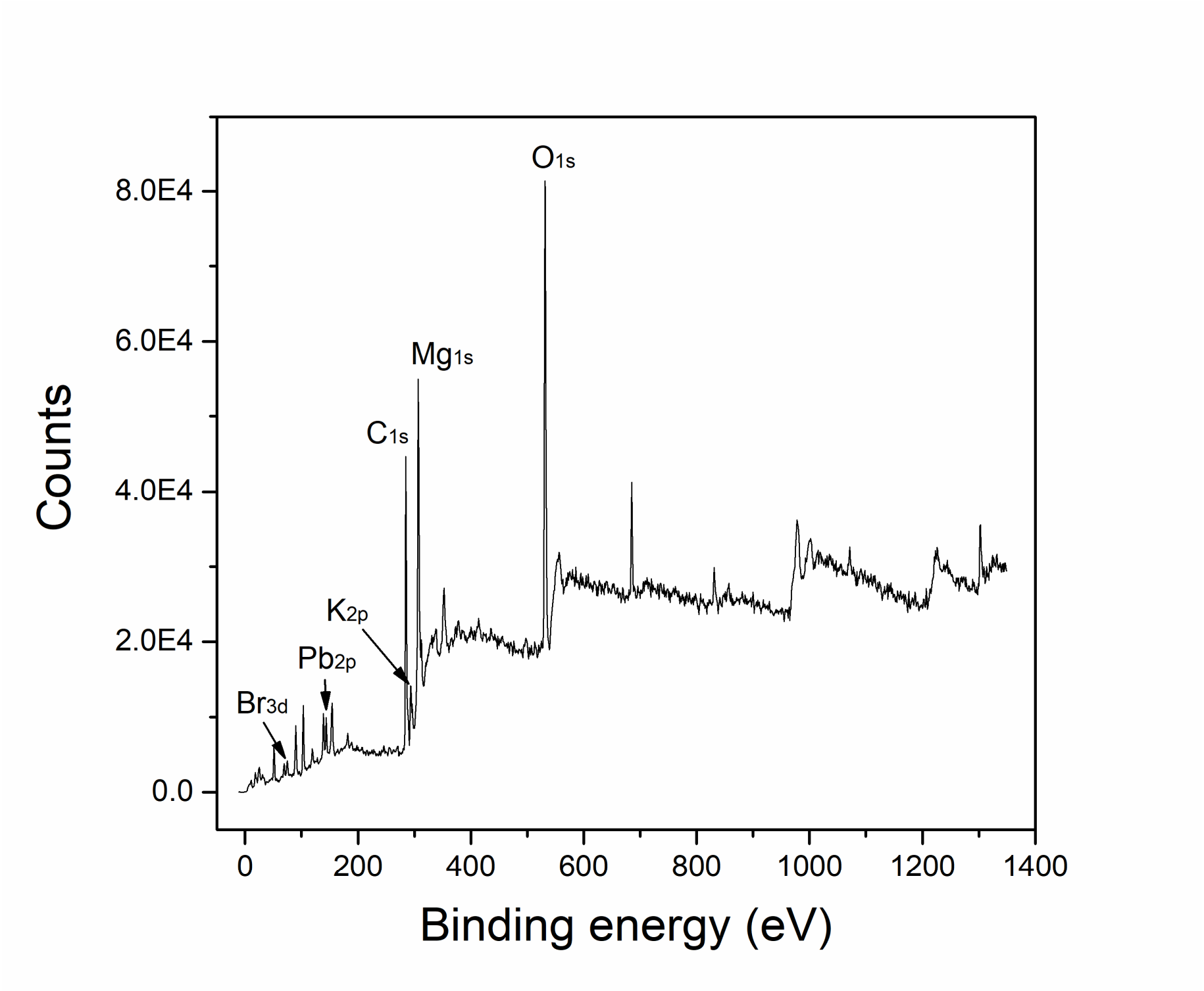

**Fig. S8. XPS spectrum of the 2D $MAPbBr_3$ perovskite on mica.**

**Fig. S9.**

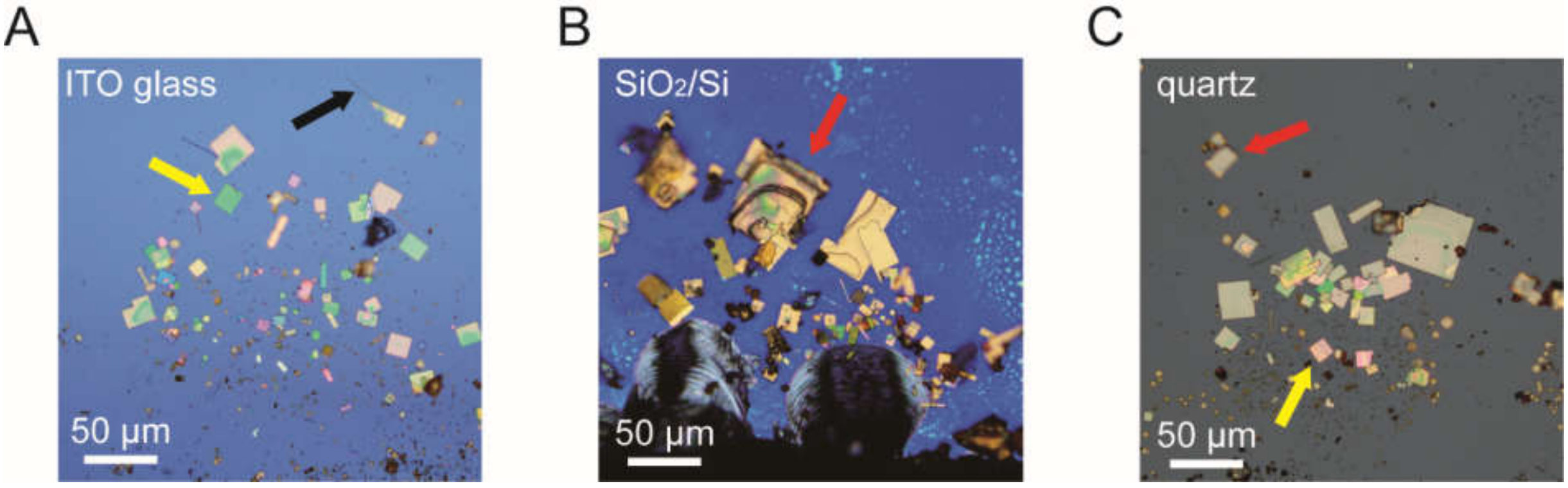

**Fig. S9. Perovskite grow on the ITO, $SiO_2$/Si and quartz substrate.** (**A**) The $MAPbBr_3$ perovskite grown on the ITO glass. (**B**) The $MAPbBr_3$ perovskite grown on the $SiO_2$/Si. (**C**) The $MAPbBr_3$ perovskite grown on the quartz. We note that the $MAPbBr_3$ perovskites prefer to crystallize in the solution with the substrates of ITO, $SiO_2$/Si and quartz, which is different from the liquid epitaxial growth on the mica surface. The $MAPbBr_3$ perovskite with different morphologies, such as thin $MAPbBr_3$ perovskite sheets (denoted by the yellow arrows), the $MAPbBr_3$ perovskite nanowires (denoted by the black arrows), the bulk $MAPbBr_3$ crystal (denoted by the red arrows), can be found.

**Fig. S10.**

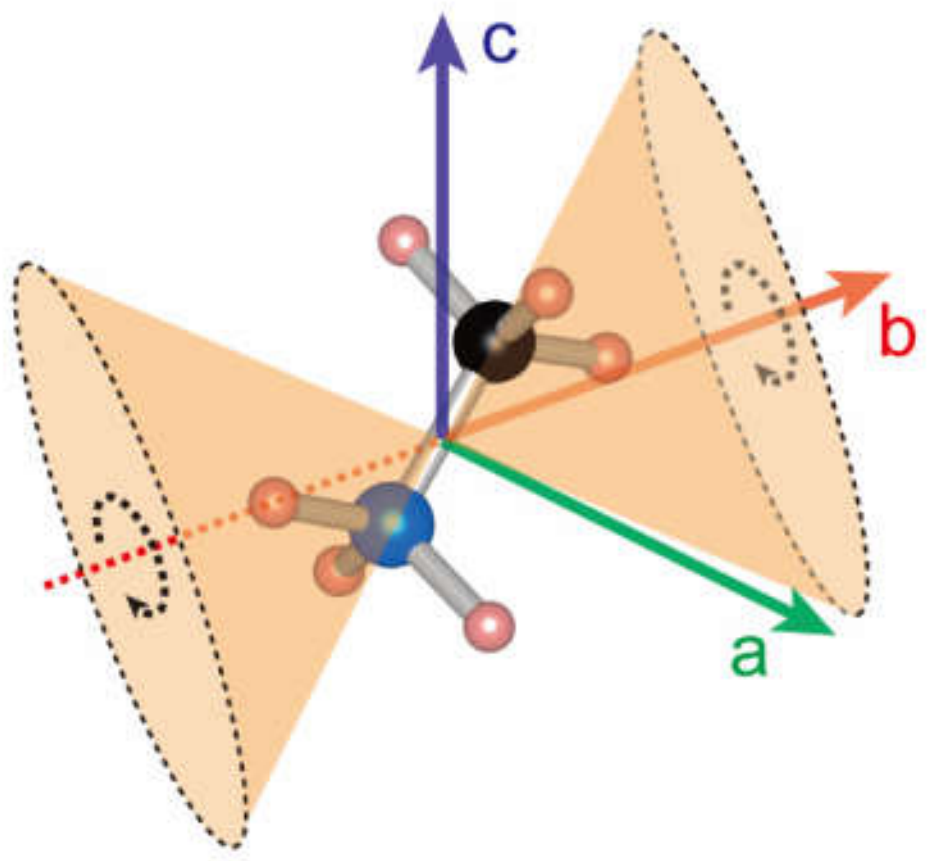

**Fig. S10. Schematic illustration of the $CH_3NH_3^-$ molecular rotation along with the (010) axis.**

**Fig. S11.**

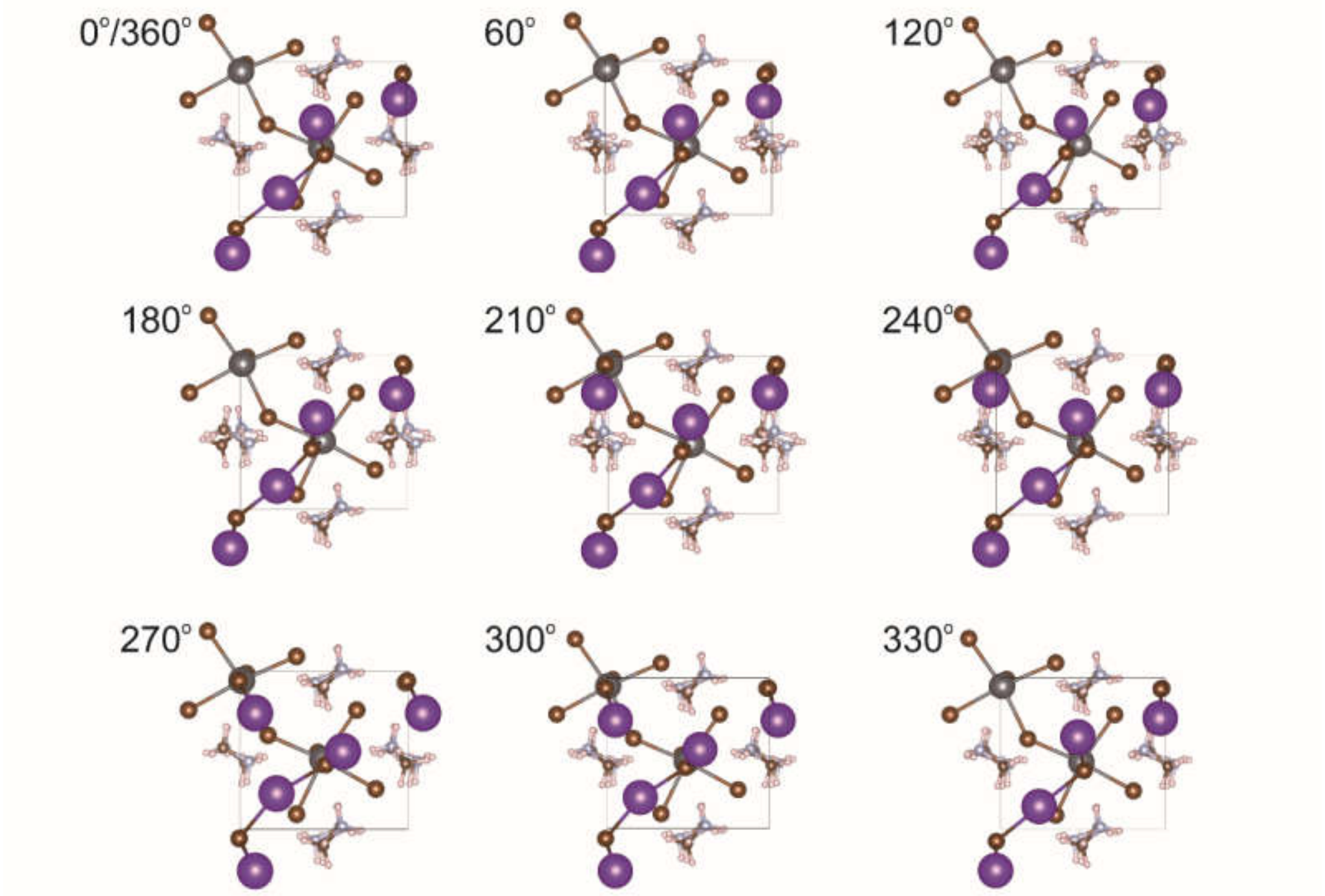

**Fig. S11. The structures of the K ions /perovskite interfaces with various spatial orientation of the $CH_3NH_3^-$ molecular.**

**Fig. S12.**

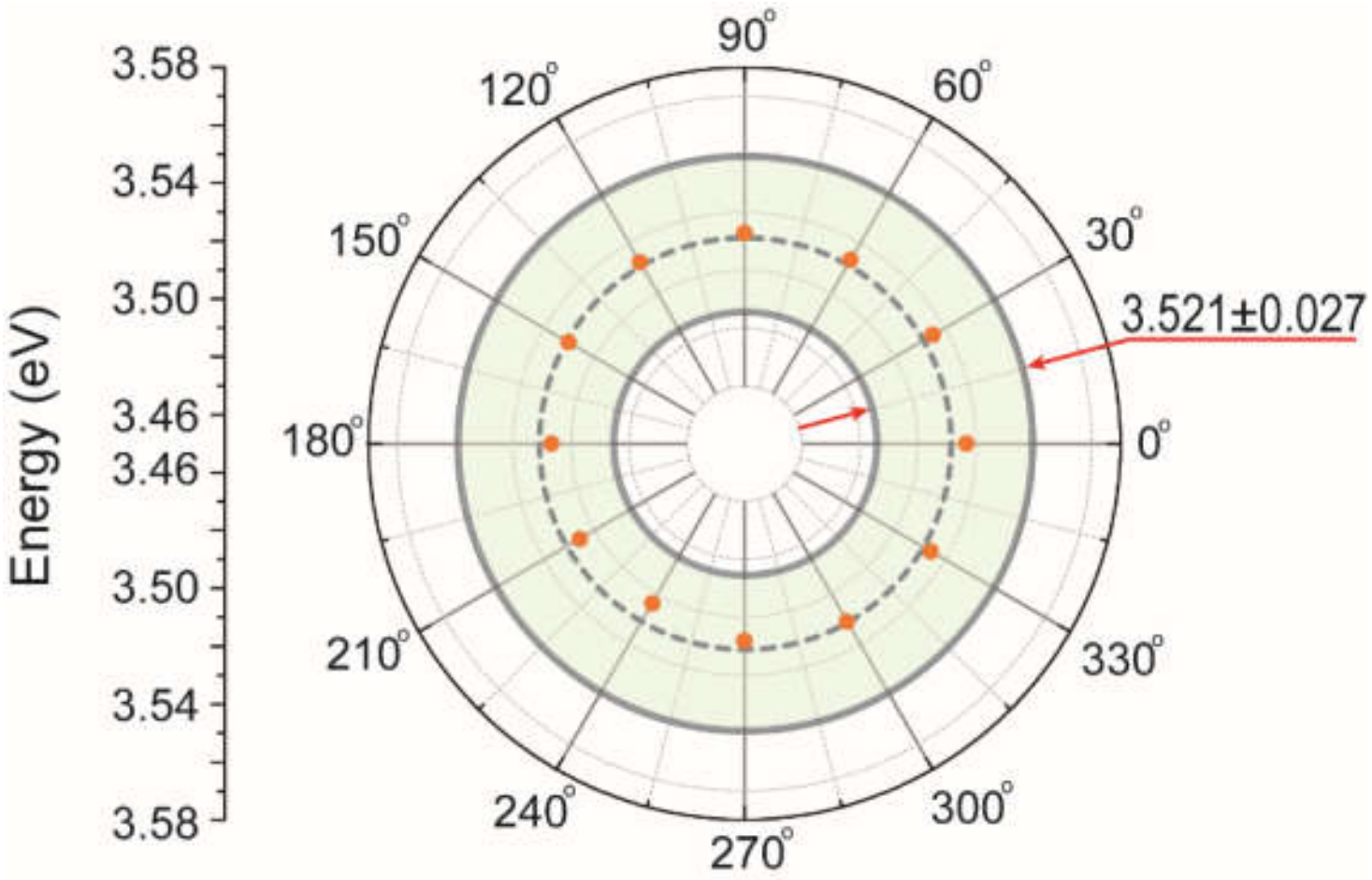

**Fig. S12. Polar coordinates of totally interface energy vs the $CH_3NH_3^-$ orientation.**

**Fig. S13.**

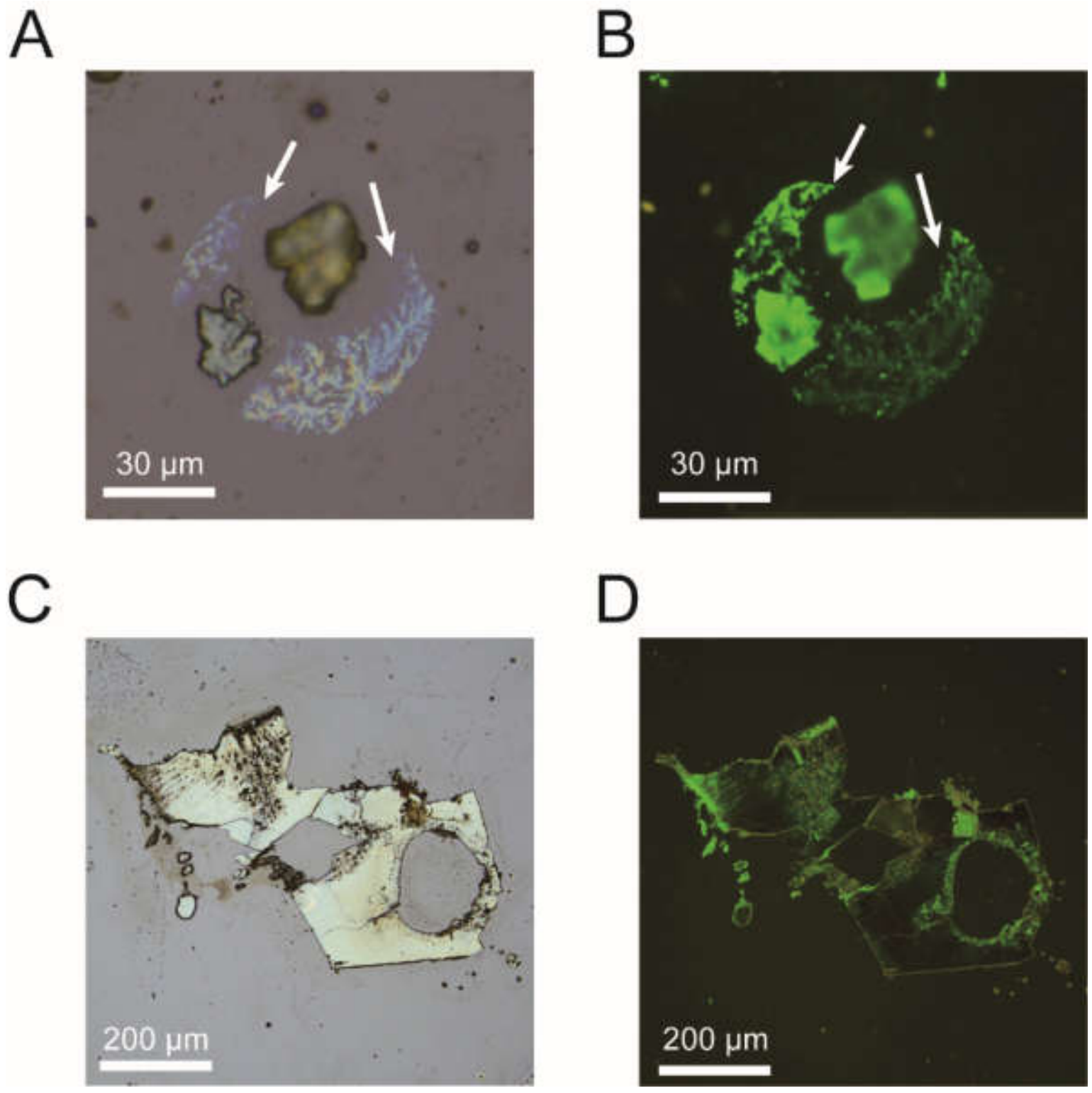

**Fig. S13. The liquid epitaxy of the 2D $MAPbBr_3$ on the KTP substrate.** (**A, B**) Optical and fluorescent images of the ultrathin $MAPbBr_3$ perovskite. (**C, D**) Optical and fluorescent images of the thick $MAPbBr_3$ perovskite.

**Fig. S14.**

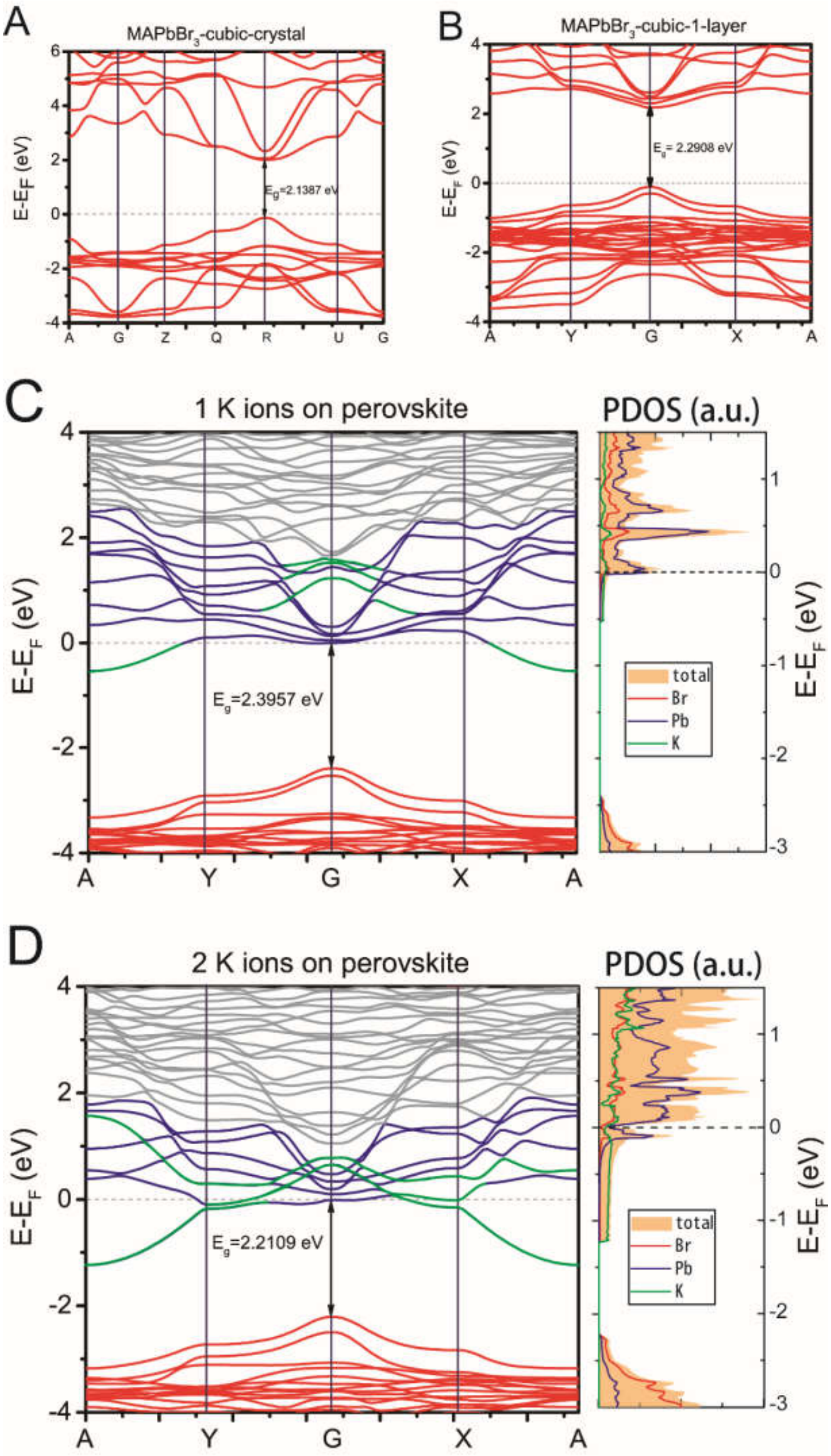

**Fig. S14. The calculated band structures of the bulk $MAPbBr_3$ perovskite, 2D $MAPbBr_3$ perovskite and the perovskite surfaces.** (**A**) The band structure of the bulk $MAPbBr_3$ perovskite, and the bandgap is estimated to be 2.14 eV. (**B**) The band structure of the 2D $MAPbBr_3$ perovskite with the thickness of single-unit cell, and the bandgap is estimated to be 2.29 eV. (**C**) The band structure of the $MAPbBr_3$ surface absorbed with one potassium ions, and the bandgap is estimated to be 2.40 eV. (**D**) The band structure of the $MAPbBr_3$ surface absorbed with two potassium ions, and the bandgap is estimated to be 2.21 eV.

**Fig. S15.**

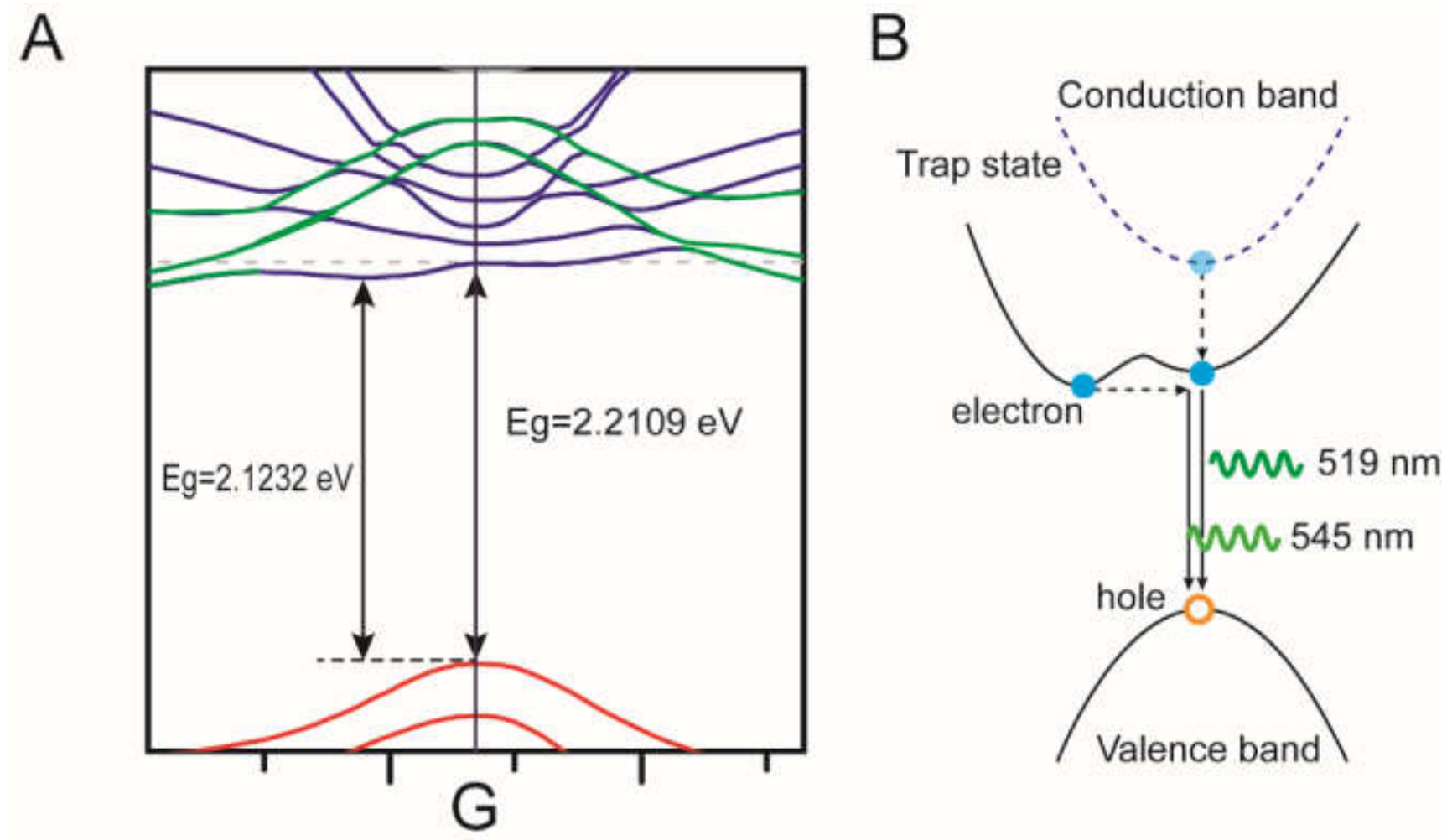

**Fig. S15. The carrier recombinations in 2D $MAPbBr_3$ perovskites.** (**A**) the band structures around at Gamma point. (**B**) Schematic illustration of the carrier recombination processes in 2D $MAPbBr_3$ perovskites.

**Table S1.**

| | $MAPbBr_3$-cubic-single layer | $MAPbBr_3$-orth-single Layer | MAPbBr3-orth-single Layer |
|---|---|---|---|
| Perovskite interface absorbed with one K ion | | | |
| Interface without K ion | -121.77247161 eV | | |
| Interface absorbed with one K ions | -121.28567324 eV | | |
| Energy of single K ion | 1.1606341 eV | | |
| Formation energy of interface | -19.819 meV/atom | | |
| Perovskite interface absorbed with two K ions | | | |
| Interface without K ion | -121.77247161 eV | -224.67433 eV | -244.6944 |
| Interface absorbed with two K ions | -120.96163574 eV | -243.75209 eV | -243.74841 |
| Energy of single K ion | 1.1606341 eV | 1.1606341 eV | 1.1606341 eV |
| Formation energy of interface | -43.155 meV/atom | -20.574 meV/atom | -20.225 meV/atom |

**Table S1. The formation energy of the K/perovskite interface calculated by DFT**

## Table S2.

| Device type | Material | Synthesis method | Responsivity ($AW^{-1}$) | Rise/Decay time (μs) | Voltage (V) | reference |
|---|---|---|---|---|---|---|
| Non-layered perovskite | $MAPbBr_3$ | Liquid epitaxy | 126 | 5.0/4.1 | 5 | This work |
| | $MAPbI_3$ | Liquid epitaxy | 20.7 | <17/- | 5 | (*38*) |
| | $MAPbI_3$ | Two-steps | 22 | $<2\times10^4/<4\times10^4$ | 4 | (*39*) |
| | $MAPbBr_3$ | Solution growth | $1.6\times10^7$ | 81/892 | 0 | (*40*) |
| Layered perovskite | $(CH_3(CH_2)_3NH_3)_2(CH_3NH_3)_{n-1}PbnI_{3n+1}$ | exfoliation | $7.19\times10^{-2}$ | -/- | 3 | (*27*) |
| | $(C_4H_9NH_3)_2(CH_3NH_3)_2Pb_3I_{10}$ | Solution synthesis | $1.278\times10^{-2}$ | $1\times10^4/7.5\times10^3$ | 30 | (*41*) |
| | $(C_4H_9NH_3)_2PbBr_4$ | Solution synthesis | $2.3\times10^{-5}$ | $3.1\times10^6/3.3\times10^6$ | 5 | (*42*) |
| | $(iBA)_2(MA)_{n-1}Pb_nI_{3n+1}$ | Solution synthesis | 0.117 | $2\times10^4/1.7\times10^5$ | 1.5 | (*43*) |
| Other 2D materials | $MoS_2$ | exfoliation | $7.5\times10^{-3}$ | $<5\times10^4/<5\times10^4$ | 1 | (*44*) |
| | $MoS_2$ | exfoliation | 880 | $4\times10^6/9\times10^6$ | 8 | (*45*) |
| | $MoS_2$ | CVD | $1.1\times10^{-3}$ | -/- | 1.5 | (*46*) |
| | WS2 | CVD | 20 | -/- | 4 | (*47*) |
| | Graphene | exfoliation | 8.61 | $>1\times10^6/>1\times10^6$ | 1 | (*48*) |

**Table S2. Device performance comparisons of the liquid epitaxy 2D $MAPbBr_3$ perovskite photodetector in this work with the reported perovskite-based photodetectors.**

**Movie S1.**
The $MAPbBr_3$ sheet growth on quartz substrate.

**Movie S2.**
The liquid epitaxy of 2D $MAPbBr_3$ perovskite on mica.